\documentclass[titlepage]{article}

\usepackage[usenames,dvipsnames,svgnames,x11names]{xcolor}

\usepackage{textcomp}

\usepackage[T1]{fontenc} %% https://tex.stackexchange.com/questions/664/why-should-i-use-usepackaget1fontenc
\usepackage[normalem]{ulem} % required for strikeout font
\usepackage{fontawesome}

\usepackage[nocompress]{cite}
\usepackage{listings}
\lstdefinelanguage{XML}
{
basicstyle=\ttfamily\footnotesize,
  morestring=[b]",
  moredelim=[s][\bfseries\color{Maroon}]{<}{\ },
  moredelim=[s][\bfseries\color{Maroon}]{</}{>},
  moredelim=[l][\bfseries\color{Maroon}]{/>},
  moredelim=[l][\bfseries\color{Maroon}]{>},
  morecomment=[s]{<?}{?>},
  morecomment=[s]{<!--}{-->},
  commentstyle=\color{gray},
  stringstyle=\color{blue},
  identifierstyle=\color{red}
}
\usepackage{moreverb}

\usepackage[nounderscore]{syntax}

\usepackage[pdftex]{graphicx}
\graphicspath{{./figures/}}
\DeclareGraphicsExtensions{.pdf}
\newcommand{\tcircle}[1]{\textcircled{\raisebox{-0.9pt}{#1}}} % circled number

\usepackage{subfig} %[caption=false,font=footnotesize,labelfont=sf,textfont=sf]
\usepackage[cmex10]{amsmath}
\usepackage{amssymb}
\usepackage{mathtools}
\usepackage{amsthm}
\usepackage{amsfonts}

\usepackage{algorithmicx}
\usepackage{algpseudocode}
\usepackage[ruled]{algorithm}
\definecolor{light-gray}{gray}{0.75}
\algrenewcommand{\algorithmiccomment}[1]{\hskip3em{{\footnotesize \textcolor{light-gray}{$\blacktriangleright$}}} #1}
\usepackage{multirow} % Multi-row tables
\usepackage{rotating} % sideways table
\usepackage{booktabs} % better lines
\usepackage{colortbl} % cell colors
\usepackage{tablefootnote} % add support for footnote in table

\usepackage[pdftex,colorlinks=true,urlcolor=blue,citecolor=blue]{hyperref}
\usepackage{xspace}

\usepackage{enumitem}
\newcommand{\para}[1]{\noindent\textbf{#1.~}}

\newcommand{\mb}{micro-batch\xspace}
\newcommand{\mbs}{micro-batches\xspace}
\newcommand{\Mb}{Micro-batch\xspace}

\newcommand{\cf}{CoFEE\xspace}

\usepackage[english]{babel}
\usepackage{blindtext}

\usepackage{authblk}

\usepackage{comment}

\begin{document}
\title{Resilient Execution of Data-triggered Applications on Edge, Fog and Cloud Resources}

\renewcommand{\thefootnote}{\fnsymbol{footnote}}

\author[1]{Prateeksha Varshney$^{*}$}
\author[2]{Shriram Ramesh$^{*}$}
\author[1]{Shayal Chhabra\footnote{Based on work done as a graduate student at the Indian Institute of Science, Bangalore, India}}
\author[3]{Aakash Khochare}
\author[3]{Yogesh Simmhan}
\affil[1]{Microsoft India R\&D Pvt. Ltd., India}
\affil[2]{Wells Fargo International Solutions Pvt. Ltd., India}
\affil[3]{Department of Computational and Data Sciences, Indian Institute of Science, Bangalore 560012, India}
\affil[ ]{\textit{prateeksha.varshney@microsoft.com}, \textit{shriramr@alum.iisc.ac.in}, \textit{shachhab@microsoft.com,}}
\affil[ ]{\textit {\{aakhochare,simmhan\}@IISc.ac.in}}

\maketitle

\begin{abstract}
Internet of Things (IoT) is leading to the pervasive availability of streaming data about the physical world, coupled with edge computing infrastructure deployed as part of smart cities and 5G rollout. These constrained, less reliable but cheap resources are complemented by fog resources that offer federated management and accelerated computing, and pay-as-you-go cloud resources. There is a lack of intuitive means to deploy application pipelines to consume such diverse streams, and to execute them reliably on edge and fog resources. We propose an innovative application model to declaratively specify queries to match streams of micro-batch data from stream sources and trigger the distributed execution of data pipelines. We also design a resilient scheduling strategy using advanced reservation on reliable fogs to guarantee dataflow completion within a deadline while minimizing the execution cost. Our detailed experiments on over 100 virtual IoT resources and for $\approx 10k$ task executions, with comparison against baseline scheduling strategies, illustrates the cost-effectiveness, resilience and scalability of our framework.
\end{abstract}

\section{Introduction}\label{sec:intro}
The Internet of Things (IoT) is leading to large-scale deployments of sensing, actuation and computing devices at the edge of the network as part of the physical infrastructure. Sensors like smart power meters, pollution monitors, and surveillance cameras are increasingly part of smart cities~\cite{simmhan:spe:2018}. Continuous analytics over observations streaming from such sensors enable efficient utility management, interventions for health and safety, and intelligent transportation~\cite{Bonomi2014}. 

Contemporary IoT and smart city applications tend to be \textit{tightly-bound} to consume specific sensor data sources. E.g., a \textit{power utility application} may monitor the \textit{kWh power load} from sensors \texttt{cds26kwh} and \texttt{cds74kwh} in the \emph{IISc campus} neighborhood, and initiate demand curtailment if the load exceeds a threshold during the peak periods of \emph{9AM--7PM}~\cite{aman:smartgridcomm:2015}. However, given the vast trove of public observation streams that change over time, such applications are more useful if they can specify ``what'' data streams they wish to consume and ``when'' they wish to consume them, rather than ``which'' specific sensor streams they should consume. E.g., rather than statically bind the power utility application above to the kWh sensor streams \texttt{cds26kwh} and \texttt{cds74kwh}, we should instead be able to state that the application should trigger for all \emph{kWh power streams} in the \textit{IISc campus spatial region} and during the \textit{9AM--7PM time period}.
So, the developer should not need to precisely know the deployed sensors and statically bind to their stream endpoints, but allow for their \emph{automated discovery and use, based on semantic needs}. There has been recent interest in using declarative queries over IoT event streams to trigger such applications~\cite{triggered-iot-2016}.

%%%%%%%%%%%%%%%%%%%%%%%%%%%%%%%%%%%%%
Another opportunity for such IoT applications is the availability of \emph{edge computing} resources as part of the infrastructure. Sensor deployments are typically connected through edge gateways such as Raspberry Pi which offer non-trivial compute resources.
Rather than use them to just move data from the sensors to the Cloud for processing, their captive compute capacity can be used to host light-weight applications~\cite{nasiri2019evaluation}. 
However, \emph{reliability} is a challenge. Further, cities are also explicitly deploying \emph{fog computing} resources with server-class or accelerated resources~\cite{amrutur:testbed:2017} to handle compute-heavy applications such as video surveillance, and allow \emph{multi-tiered management} of edge devices~\cite{he-2017}. Edge and fog are \emph{cheap or free captive resources}, and can serve as the first-line of computing for smart city applications. However, their resource capacities need to be supplemented on-demand by pay-as-you-go cloud resources~\cite{Buyya:2018:manifesto}. At the same time, \emph{platforms} to ease the development and robust deployment of applications on edge, fog and cloud are still evolving~\cite{bittencourt2017mobility}.

%%%%%%%%%%%%%%%%%%%%%%%%%%%%%%%%%%%%%
In this paper, we leverage the twin-opportunities of \emph{ubiquitous data streams} and \emph{edge and fog computing resources} in smart cities, to ease the specification of declaratively-triggered applications over such streams, and execute them reliably on edge, fog and cloud resources. We make these contributions:

\begin{enumerate}
	\item We propose a novel \emph{trigger-based model} for dataflow execution based on \emph{declarative queries} specified over the attributes of \mb streams generated from wide-area sensor sources (Sec.~\ref{sec:sys-data-model},~\ref{sec:app-model}).
	\item We develop a unique \emph{resilient decentralized scheduling strategy} for \textit{dynamically instantiated} dataflows on unreliable edges, and reliable fog and cloud resources, using \textit{advanced slot reservation} on the fog, while meeting the dataflow deadline and minimizing the cost (Sec.~\ref{sec:schedule}).
	\item We offer \emph{detailed experiments on 100+ emulated edge and fog resources} for realistic dataflows with $10k$ task executions to validate the lower cost, higher resilience and scalability of our approach relative to baselines (Sec.~\ref{sec:results}).
\end{enumerate}
In addition, we also discuss related works (Sec.~\ref{sec:related}) and offer our conclusions (Sec.~\ref{sec:conclude}).

%% #############################################################
\section{System and Data Model} \label{sec:sys-data-model}

%%%%%%%%%%%%%%%%%%%%%%%%%%%%%%%%%%%
\subsection{System Model} \label{sec:sys-model}
Edge and fog are emerging resource abstractions that complement the well-established cloud computing~\cite{Buyya:2018:manifesto}. Given the diverse definitions for the edge and fog paradigms, we state our assumptions on their characteristics~\cite{varshney2017icfec}.

\emph{Edge resources} are distributed across a metropolitan area network (MAN) using wired, cellular or \textit{ad hoc} network connectivity. They are often co-located with sensors that serve as sources of data streams. Edges have low-end computing and memory capacity, e.g., a Raspberry Pi with a multi-core ARM processor and 1-2~GB RAM. These devices may be \emph{unreliable}, e.g., due to mobility, intermittent network, on-field failures, energy constraints, etc. Typically, these are captive and free resources as part of a city infrastructure, but finite in number. They may use containers for application sandboxing.

\emph{Fog resources} are also distributed across a MAN, but \emph{reliable} and connected to a high-speed broadband or cellular network. They offer a workstation or server-class performance and optionally have accelerators. They are also captive, finite and as cheap as or costlier than the edge resources. They may use containers or hypervisors as the application environment.
	
\emph{Public cloud resources} at data-centers are accessible over the Wide Area Network (WAN). They offer unlimited on-demand access to \emph{reliable} virtual machines (VM) with server-grade CPUs and diverse capacities. Their pay-as-you-go pricing may be costlier or cheaper than fog resources.

\para{Management} Since Edge devices are constrained, we assume that the edge and fog resources are organized hierarchically to ease their management. Specifically, the resource and application management for each edge is done by its \emph{parent Fog}, and a parent fog and all its edge children form a \emph{fog partition}~\cite{he-2017}. This grouping may be based on network or spatial proximity, or other organizational domains. Any interaction with an edge, say to schedule a task or to transfer data, is only through its parent fog which serves as a gateway. We expect the network performance between edge--parent fog to be high, and from fog--fog and fog--cloud to be lower.

\para{Definitions} Let $r^E_i \in \mathbb{R}^E$, $r^F_j \in \mathbb{R}^F$ and $r^C_k \in \mathbb{R}^C$ be the set of edge, fog and cloud resources, respectively, with $\mathbb{R} =  \mathbb{R}^E \cup  \mathbb{R}^F  \cup \mathbb{R}^C$ as the set of all resources in the system. $C(r_i^F) \subset \mathbb{R}^E$ is the set of \textit{edge children} for a parent fog $r_i^F$, at a given time. The set $\mathbb{R}^E$ can vary over time.  $\epsilon$ is the \emph{incremental time unit of billing} for a resource with a \emph{price function} $\pi(r_x)$ that gives the fixed-price for the resource $r_x$ for $\epsilon$ time units.
The \emph{performance scaling}, or compute speed, of a resource relative to a baseline (slowest) resource $r_0$ is given as the function $\rho(r_x)$, with $\rho(r_0)=1$.
So the time taken to execute a task on $r_x$ will be $\frac{1}{\rho(r_x)}$ of the time taken for the task on $r_0$. This can be workload-specific. The network bandwidth between resources $r_x$ and $r_y$ is given by $\beta_{xy}$ and the latency by $\lambda_{xy}$; $\beta=0$ for resources that cannot reach each other. $\phi_{xy}$ is the monetary cost for unit data transfer between two resources. These coefficients allow the scheduler to take performance and price-aware decisions.

%%%%%%%%%%%%%%%%%%%%%%%%%%%%%%%%%%%
\subsection{Streaming \Mb Data Model} 
\label{sec:data-model}

The primary input source to our applications is data streaming from sensors, typically connected to an edge and sometimes a fog resource. Rather than target real-time event-based streaming applications, we instead consider the streaming \emph{\mb model}~\cite{de2018distributed} used by systems like Spark Streaming~\cite{nasiri2019evaluation}. Here, a time or a count window of time-series observations is accumulated into a \emph{\mb}, which forms the logical unit of execution, data movement and storage. These \mbs themselves may be constantly generated from the sensor stream, forming a \emph{stream of \mbs}. This model offers better \emph{throughput} since the data movement, scheduling and execution overheads are amortized across all events in that \mb, while bounding the latency overheads to the \mb size. The mechanism of acquiring data from the streams and forming \mbs is outside the scope of this paper~\cite{de2018distributed,ravindra2017mathbb}.

Besides the \emph{content} formed from the window of observations, a \mb also has several \emph{spatio-temporal and domain attributes}.  This metatata is crucial for automated discovery of such data sources and declarative binding of applications to them. A \mb $\mu_{i}$ is defined as the tuple: 
\[
\mu_i = \langle~ id, sid,  \langle t_b,t_e \rangle,  \langle lat,long \rangle,  \langle key,val \rangle^*, size, content ~\rangle
\]
where $id$ uniquely identifies the \mb, $sid$ identifies the source from which it was generated, $\langle t_b,t_e \rangle$ are the begin and end timestamps of the contents, $\langle lat,long \rangle$ are the spatial context for the data source, $\langle key,val \rangle^*$ are a set of domain-specific key--value metadata, $size$ is the length in bytes of the content, and $content$ has the actual {\mb}ed data.

E.g., the following \mb represents $8$ observations accumulated from a temperature sensor \texttt{cds26temp} present in a campus IoT deployment on \emph{Nov 15, 2021}.~\cite{simmhan:spe:2018}.

{\small
\begin{lstlisting}%[aboveskip=0pt]
$\langle$9cdfa00dc01d, cds26temp,
$\langle$2021-11-15T09:00:00, 2021-11-15T09:05:00$\rangle$, 
$\langle$13.0165, 77.5706$\rangle$, [$\langle$units, C$\rangle$, $\langle$err, 0.05$\rangle$], 39, 
[27.5,27.5,27.6,27.7,27.7,27.7,27.6,27.7]$\rangle$
\end{lstlisting}
}

%% #############################################################
\section{Data-Triggered Application Model} \label{sec:app-model}
\begin{figure}[t]
	\centering
	\includegraphics[width=0.9\columnwidth]{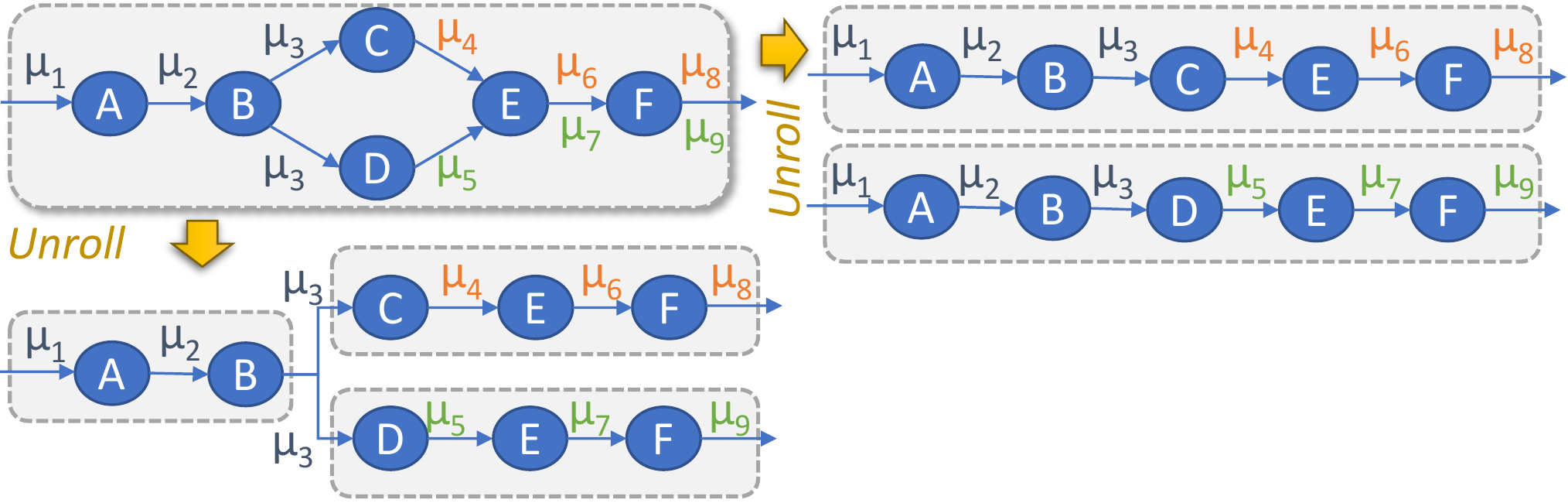}
	\caption{Unrolling of input DAG (top-left) into linear pipelines (top-right) or nested linear pipelines (bottom-left).}
	\label{fig:dag-unrolling}
\end{figure}

We propose a \emph{deadline-driven dataflow model} for application composition with a novel \emph{triggering} of its execution based on \textit{declarative matching} of user queries with distributed \mbs. We first specify the dataflow model followed by the query-based triggering approach.

%%%%%%%%%%%%%%%%%%%%%%%%%%%%%%%%%%%
\subsection{Application Definition}
\label{sec:app:defn}

We use the common \emph{Directed Acyclic Graph (DAG)} as our application model. Users define their application as a DAG, $\mathcal{D} = (\mathbb{T}, \mathbb{E})$, with a set of vertex tasks $\tau_j \in \mathbb{T}$ and their dataflow edges $\langle \tau_i, \tau_j \rangle \in \mathbb{E} \subseteq \mathbb{T} \times \mathbb{T}$. Each task $\tau_j$ consumes a single \mb as input, and produces one \mb as output on completion.
An edge $\langle \tau_i, \tau_j \rangle$ indicates a data dependency between tasks -- an output \mb generated by the execution of $\tau_i$ will be used as input for the execution of task $\tau_j$. Users also provide the scheduler with a \emph{baseline execution time}, $\theta_j$, for each task $\tau_j$ when executed on a base resource $r_0$, and the \emph{expected size} of the output \mb from the task, $\alpha_j$, based on prior benchmarking.

While a DAG offers substantial flexibility, most practical applications tend to be simple linear pipelines with stateless tasks. This is seen in the growing popularity of microservice tasks on the cloud defined as short-running and stateless Functions-as-a-Service (FaaS) and composed as a linear chain of event-driven computing, and executed using patterns such as Saga~\cite{foxWorkshop}.
As a result, we make several practical simplifying assumptions. Tasks are \emph{stateless} and a \mb is its unit of execution on a single resource. Aggregation can be done within events in a \mb but not across. In case a task executing a \mb fails due to a resource failure, it is \emph{re-executed} for the same \mb on another resource. 

The dataflow uses \emph{interleave semantics}, i.e., if edges from two upstream tasks are incident on a downstream task, the latter will independently execute once for each output \mb from the two upstream tasks. This downstream path-independence lets us easily unroll the DAG into linear chains, converting a task-parallel to a data-parallel execution. E.g., in Fig.~\ref{fig:dag-unrolling}, the task $E$ of the input DAG in the top-left executes twice for inputs \textcolor{orange}{$\mu_4$} and \textcolor{ForestGreen}{$\mu_5$} and the downstream execution path for their outputs \mb \textcolor{orange}{$\mu_6$} and \textcolor{ForestGreen}{$\mu_7$} are independent. This in equivalent to unrolling the DAG into two \emph{sequential pipelines} shown on the top-right. A future optimization can split the common precursor sub-chains hierarchically (Fig.~\ref{fig:dag-unrolling}, bottom-left) to avoid duplicate execution of tasks, e.g., $A$ and $B$.
This approach eases the execution model while limiting penalties for DAGs without too many branches.

The user also specifies a \emph{deadline} $\delta$ for the execution of the DAG for a single input \mb from its time of arrival. So all causal \mbs generated for this input \mb should be executed by the DAG tasks within this deadline, i.e., all linear pipelines should complete by this deadline.

%%%%%%%%%%%%%%%%%%%%%%%%%%%%%%%%%%%

\subsection{Application Triggering} \label{sec:app:trig}

Each dataflow's runtime execution is in the context of a single \mb. A key challenge is to identify \mb streams on which the dataflow should execute. Binding the dataflow to pre-defined stream endpoints requires \textit{a priori} knowledge of all streams, which is difficult in an evolving IoT environment. Filtering \mbs with certain attributes as part of an initial task of a DAG will still pay the cost for moving the \mb to the task even if it does not match. Similarly, publishing the \mb to a central event broker for filtering causes unnecessary data movement on the WAN.

Instead, we propose a novel \emph{declarative model} for specifying the characteristics of the input \mb, based on its metadata, for a dataflow. Our runtime then automatically matches the generated \mbs from \emph{various streams across the edge and fog layers} to trigger a DAG execution on any \mb. Since the \mb attributes include spatial, temporal, sensor and domain attributes, this allows for fairly \emph{complex patterns} to be defined easily by the user and matched at the \emph{granularity of each \mb}. The application also implicitly \emph{adapts to streams} entering and exiting as they are co-located with the edge resources. Users can always define a simple query that matches the sensor ID if they wish to statically bind their application to a specific sensor's stream. The matching is \textit{light-weight, distributed across fogs and avoids data movement}, with a throughput of $1000$s of \mb/sec on just an ARM-based fog.

As part of the DAG submission, users specify a declarative \emph{filter query} $\mathcal{F} = \langle f_s, f_t, f_d \rangle$
where $f_s, f_t$ and $f_d$ are query predicates defined on the \emph{spatial coordinates, time range} and \emph{domain properties or sensor id} metadata of each \mb, respectively. 
This filter is active until the DAG is undeployed. The filters registered for the deployed DAGs are matched by the \textit{query-matching engine}, as discussed next, against all available streams on all the edge and fog resources. So, the users focus on \emph{what} to match rather than \emph{where} the source is present. 
Any \mb matching the filter triggers the execution of an instance of the filter's DAG. 
The execution of each input \mb for a DAG is independent of other matched \mbs. The same \mb can be matched by filters for different DAGs, triggering all of them. There is no ordering guarantee for different \mbs for the same DAG, making \mb execution independent of their generation time. This allows \mbs to make use of \emph{task, data and pipeline parallelism} to the full extent. The use of \mbs combined with low-latency matching and immediate deadline-aware scheduling for execution balances the throughput and the execution latency, while allowing maximal utilization of edge and fog resources.

%%%%%%%%%%%%%%%%%%%%%%%%%%%%%%%%%%%%%%%%%%%%%%%%%

\subsection{Application Execution Engine}
\label{sec:app:exec}

\begin{figure}[t]
	\centering
	\includegraphics[width=0.99\columnwidth]{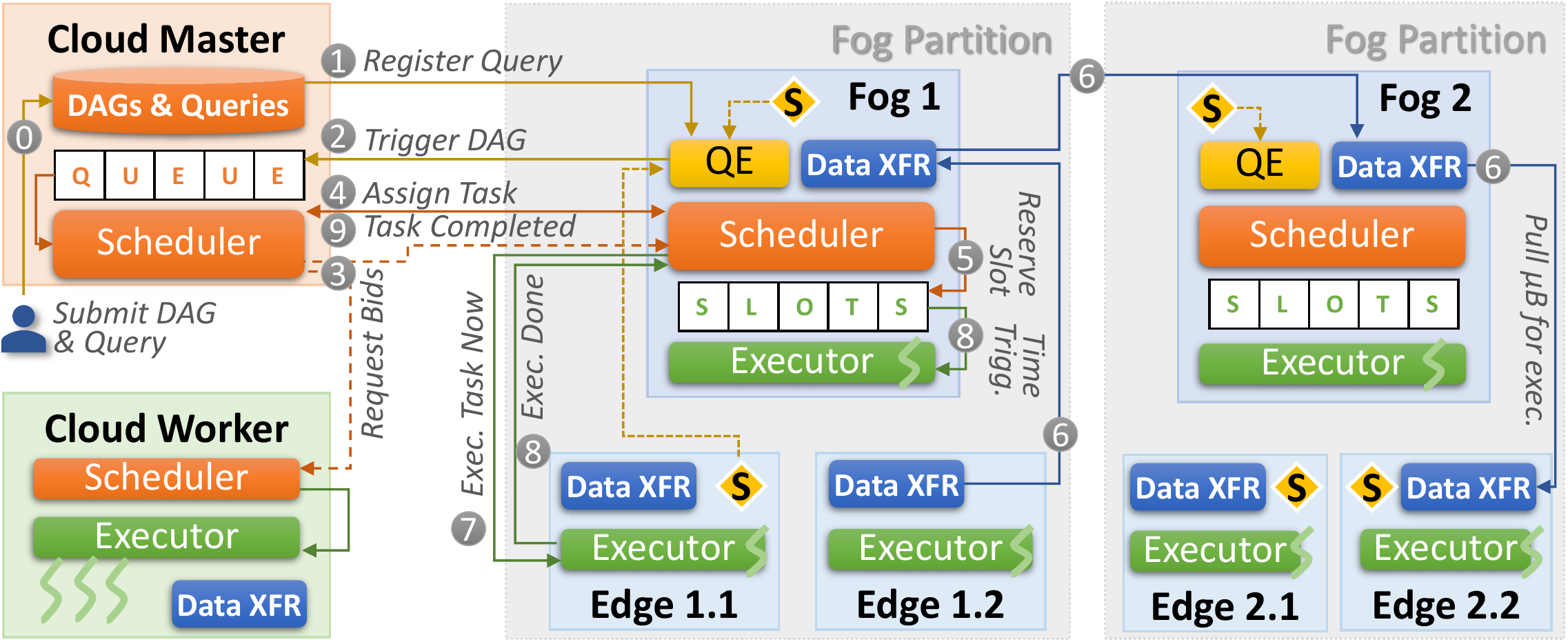}
	\caption{Architecture of \cf on Edge, Fog and Cloud}
	\label{fig:arch}
\end{figure}

We have designed the \underline{\bf C}l\underline{\bf o}ud, \underline{\bf F}og and \underline{\bf E}dge \underline{\bf E}xecution (\cf) Engine for orchestrating the filter matching, DAG triggering, schedule planning, and resilient execution of the user provided dataflows on edge, fog and cloud resources. Here, we describe its execution model; in the next section, we propose heuristics for their deadline-aware and resilient scheduling.
The architecture (Fig.~\ref{fig:arch}) consists of edge, fog and cloud resources \textit{(Workers)} that are available for task execution. We also have a separate cloud VM that runs a \emph{Master service} for distributed coordination and holding light-weight state.

\subsubsection{Federated Resource Discovery}
The cloud and fog workers initially register with the Master. Edge resources bind with a parent fog when they join the system. The parent is chosen by querying the Master based on proximity, domain rules or prior know\-ledge. The fog schedules tasks and transfers data for edges in its partition. When an edge goes offline, it may explicitly inform its parent fog or implicitly drop off by failing to send heartbeats to it. This \emph{federation} helps scale to a large number of edge and fog resources. 

\subsubsection{DAG Registration and Triggering}
\emph{Sensor services} ({\bf S} in Fig.~\ref{fig:arch}) generating \mb streams may be present on any edge or fog. \emph{DAGs} are submitted by the user to the Master along with its \emph{filter query} and \emph{deadline} (Fig.~\ref{fig:arch}~\tcircle{0}). The Master stores these DAG definitions locally, and registers the query with a light-weight \emph{query-matching engine (QE)}  running on each fog (\tcircle{1}). The sensors on the edge or fog stream the \emph{metadata attributes} for \mbs they generate as events to the QE on the parent fog. Each event is $\approx 100~bytes$ in size. The QE matches the metadata events from sensors in this partition against the registered queries.
We currently support equality matching for the domain attribute, and equality, intersecting or contains matching on the spatial and temporal ranges, and do an \texttt{AND} on the matching predicates.
If a \mb's metadata matches a filter, it \emph{triggers} the execution of the filter's DAG on that \mb with the Master (\tcircle{2}). A \mb may match multiple DAGs and trigger multiple executions.
By leveraging the fog for querying, we load balance the query cost across partitions and avoid the Master being a bottleneck.

\subsubsection{DAG Scheduling}
We discuss finer details of the scheduling heuristics in Sec.~\ref{sec:schedule}, but review the approach here. When a fog matches a \mb on its partition and triggers a DAG (\tcircle{2}), the Master puts the DAG and the \mb ID on its scheduling \emph{queue}. The Master then \emph{unrolls} the DAG into linear pipelines and apportions the DAG's deadline as \textit{sub-deadlines} for the tasks in each pipeline. The source task for each pipeline is first scheduled, followed by each subsequent task after the completion of its predecessor. 

Scheduling a task follows a \emph{inquiry, bid and select cycle}. The Master contacts a subset of under-loaded and cheap fogs with the task's baseline execution time, the sub-deadline and location of the input \mb (\tcircle{3}). Each Fog's \emph{Scheduler} responds with a bid if it can execute the task within the deadline on workers (edges or parent fog) in its partition, and returns the estimated cost based on its resource pricing. The Master also gets a bid from the on-demand Cloud worker. It then selects the cheapest fog partition or the cloud worker which can successfully execute the task, and assigns the task to it (\tcircle{4}). When a task completes, the Master is informed of its output \mb ID (\tcircle{8}). The Master then schedules the next task in the pipeline using a similar cycle. The scheduler logic of the Master is stateless and can scale concurrently, while its state and the queue can be replicated for resilience.

\subsubsection{Data Transfer and Execution}
A task that is scheduled on a fog partition for execution may run on an edge or the fog, based on the bid. If the input \mb to the task is on a different resource, we use \emph{data transfer services} running on each resource to pull the input \mb and stage it locally on the target resource before execution (\tcircle{6}). The data transfer services of fogs and cloud workers can talk to each other since they are usually on public networks, but edges on private networks use their parent fog's data service to exchange data. E.g., in Fig.~\ref{fig:arch}, \emph{Edge 2.2} pulls an input \mb for a local task from \emph{Edge 1.2} through \emph{Fog 2} and \emph{Fog 1}.

\subsubsection{Resilience} Without loss of generality, only one task executes on an edge or a fog at a time due to their limited resources, but the execution on a cloud worker can be concurrent as its capacity and pricing scales linearly. A task executes on an edge's \emph{Executor} service \textit{immediately} after the fog assigns it (\tcircle{7}). But an execution on the fog may happen in the \emph{future} (but before its sub-deadline) based on an \emph{advanced slot reservation} decided by the fog (\tcircle{5}, \tcircle{8}). The task completion is made tolerant to edge failures through slot reservation by the parent fog on itself for \emph{fail-over and re-execution} (\tcircle{8}), and by \emph{caching} the task's input \mb (\tcircle{6}). Once a task completes, the output \mb is stored on the executing resource, the parent fog informs the Master of the completion, and passes it the output \mb ID to trigger the next task in the pipeline (\tcircle{9}). These are discussed next in Sec.~\ref{sec:schedule}.

\subsubsection{Implementation}
The \cf Engine implements all scheduling and runtime logic as \emph{micro-services} in \emph{Python} using \emph{Google RPC} and \emph{Protobuf}. This makes them light-weight and portable to execute even on low-end edge devices. Specifically, we have micro-services for the QE, the scheduler, the data transfer and the task executor, which variously run on edge, fog and cloud. All edge and fog services run within a \emph{container} environment for sandboxing. For simplicity, the task binaries are pre-installed as part of container or VM image creation.

%%%%%%%%%%%%%%%%%%%%%%%%%%%%%%%%%%%%%%%%%%%
\section{Deadline-aware Resilient Scheduling} \label{sec:schedule}
The scheduling heuristic that we propose optimizes for \emph{(1) reducing the monetary cost of executing the DAG, (2) while meeting its deadline, and (3) being resilient to Edge failures}.
Building upon the high-level objectives above, the specific scheduling strategies are discussed in detail next.

\subsection{Inquiry and Selection of Resource upon Trigger}\label{sec:schedule:trigger}
\subsubsection{Trigger} When a \mb from an edge or fog sensor stream matches a filter query registered with the QE on the parent fog, it triggers the DAG on the Master with the \mb ID. The Master unrolls the corresponding DAG into multiple sequential pipelines of tasks. 

The deadline for the entire DAG also applies to each pipeline. The \emph{sub-deadline for each task} in the pipeline is calculated by distributing the DAG's deadline $\delta$ among the tasks, in proportion to the task's execution time contribution to the pipeline~\cite{Abrishami:2013:DWS:2388122.2388265}.
For a pipeline $\mathbb{P}$ having tasks $[\tau_1, \tau_2, ..., \tau_p]$, the sub-deadline $\sigma_i$ of a task $\tau_i$ is:
$ \sigma_i = \frac{\theta_i}{\sum_{j=1}^{p} \theta_{j}} \times \delta $, where $\theta_j$ is the baseline execution time for a task $\tau_j$.

\subsubsection{Inquiry} On a trigger, the Master initiates the \emph{inquiry phase} to find the cheapest fog partitions available to execute the \textit{source task} for a pipeline. Since there may be $10$--$100s$ of fog partitions, broadcasting the inquiry to all fogs has a high overhead. But sending the request to more fogs will help discover the cheapest viable one. To balance between these, the Master sends an inquiry request to the \textit{top-n} least-loaded and cheapest fog partitions. For this, each fog periodically reports its \emph{top-k} longest free slots available to the Master. As discussed next, these free slots are a good proxy for the load on the fog partition since tasks running on the edge and scheduled on the fog will reserve these slots. From these, the Master picks fogs with free slots long enough to execute the task within its sub-deadline. Among them, it picks the \textit{top-n} with the lowest resource cost for the parent fog. This increases the chance of getting a viable bid, and also prefers cheaper fogs.

The \emph{inquiry request} is sent to these fogs with details of the source task $\tau_0$ and input \mb, $\langle \tau_0, \theta_0, \sigma_0, \mu_x, \alpha_x, r_x \rangle$, which provide the task's ID, baseline execution time and sub-deadline, and the \mb's ID, size and the resource it is present on, based on the initial trigger. The Master waits for a pre-defined timeout period, $t_{inq}$, to receive the \emph{bid} responses from these $n$ Fogs.

\subsubsection{Selection} After the timeout period expires or when all $n$ fogs respond with a bid, whichever is earlier, the Master picks the viable bid that can complete the task within its deadline and has the cheapest cost, $\kappa_{min} = \min_n(\kappa)$. It sends that fog an \emph{accept bid} message, asking it to proceed with the task execution; it also sends a \emph{reject bid} message to the other viable fogs.

It is possible that none of the fog partitions submit a viable bid for a task inquiry because their edges or fog are busy until the task deadline. We implicitly always include a bid from the Cloud worker in our selection process since it is always available, on-demand. We first estimate if the Cloud VM can complete the task within the sub-deadline, considering the \mb transfer time and the task execution time, and if so, we estimate the cost for execution to form a bid. This is considered when selecting the cheapest bid from the fogs. So even if the resource cost for the cloud is cheaper than the fog, our scheduling heuristics will work. If none of the Fog partitions submit a bid and if the Cloud VM cannot finish before the deadline, this task and hence pipeline has failed.

\subsection{Bidding by the Fog for an Inquiry}\label{sec:schedule:bid}
A fog  $r_i^F$ that receives an inquiry $\langle \tau_y, \theta_y, \sigma_y, \mu_x, \alpha_x, r_x \rangle$ checks if it can execute the task $\tau_y$ on the input \mb $\mu_x$ within the sub-deadline on any of its edges, $r_j^E \in C(r_i^F)$, or on itself. For this, it first checks these conditions on each idle edge in its partition not currently executing a task:
\begin{center}
% \vspace{-0.05in}
$ \omega_j + \frac{\theta_y}{\rho(r^F_i)} ~\le~ \sigma_y \qquad \text{where~~} \omega_j = t_{inq} + d_{xj} + \frac{\theta_y}{\rho(r^E_j)}$
\end{center}
This tests if the time taken for: the inquiry and bid phase to complete, the data transfer $d_{xj}$ from the current \mb location at $r_x$, the task execution on the edge and also the task re-execution on the fog parent \emph{if} the edge fails, all together fall within the sub-deadline $\sigma_y$. Here, $\omega_j$ gives the \emph{latest completion time} on the edge; if the edge fails to complete the task by this time, we will re-execute it on the fog. We use estimates of the inter-resource bandwidths $\beta_{xi}$ and $\beta_{ij}$ between the resource $r_x$ hosting the \mb to this parent fog $r^F_i$ and from this parent fog to the local edge $r^E_j$, and similarly their latencies $\lambda_{xi}$ and $\lambda_{ij}$, combined with the \mb size $\alpha_x$, to estimate the incoming \mb transfer time as $d_{xj} = \big( \lambda_{xi} + \frac{\alpha_x}{\beta_{xi}} \big) + \big( \lambda_{ji} + \frac{\alpha_x}{\beta_{ij}} \big)$. The task execution time is scaled by the relevant processing speed $\rho(\cdot)$ of the resources.

For every edge device $\widehat{r}_j^E$ which satisfies the above condition, the fog computes the \emph{expected maximum cost}, $\kappa_j$, for running the task. This is the cost of executing the task on that edge, and the \emph{probabilistic cost} of having to execute the task on the fog if the edge fails to complete it by the deadline.
\begin{center}
$\kappa_j = \big( k^E + \big\lceil \frac{\theta_y}{\rho(\widehat{r}_j^E) \cdot \epsilon} \big\rceil \cdot \pi(\widehat{r}_j^E) \big) + \big( k^F + P(\widehat{r}_j^E) \cdot \big\lceil \frac{\theta_y}{\rho(r_i^F)\cdot\epsilon} \big\rceil \cdot \pi(r_i^F)  \big)$
\end{center}
The first part is the cost on the edge, including \emph{monetary cost for data transfer}, $k^E = \alpha_x \times (\phi_{xi} + \phi_{ij})$, and the \emph{task execution cost} based on the execution time, the billing increment $\epsilon$, and price of the edge $\pi(\cdot)$.
The second part is the probabilistic cost on the fog, which is similar, but includes the probability of failure of the edge, given by $P(\widehat{r}_j^E)$. This probability can be a fixed value provided for the edge, or calculated from its \textit{Mean Time Between Failure (MTBF)}.
$k^F = \alpha_x \times \phi_{ji}$ is the cost to copy the input \mb to the parent fog from the edge -- 
we always cache a copy of the input \mb on the parent fog to allow for re-execution if the edge fails. These \mb transfer times and costs are adjusted accordingly if it is already present in the local fog partition.

The fog returns the candidate edge with the cheapest expected maximum cost back to the Master. However, we still need to check if the parent fog has the capacity to re-execute the task if the edge fails. For this, we use a slot reservation strategy discussed next. If none of the edges can meet the deadline, or if the fog cannot reserve a future re-execution slot, we check if the fog can run the task directly and meet the sub-deadline; if so, we return its cost to the Master, and if not we return an empty bid to the Master.

\subsection{Advanced Slot Reservation for Fogs}\label{sec:schedule:reserve}
It may be possible to delay executing a task after its submission and still complete it within its sub-deadline. We use this intuition to enable re-execution of a task on the parent fog if it fails on an edge. As concurrent tasks can execute on different edges of the partition and may fail simultaneously, we must ensure that all of them can re-execute on the parent fog within their sub-deadline. Further, tasks may also be directly run on the fog. So we need to carefully plan the tasks scheduled on the fog to avoid over-allocation of its resources and guarantee the completion of tasks assigned to its partition.

\begin{figure}[t]
	\centering
	\includegraphics[width=0.95\columnwidth]{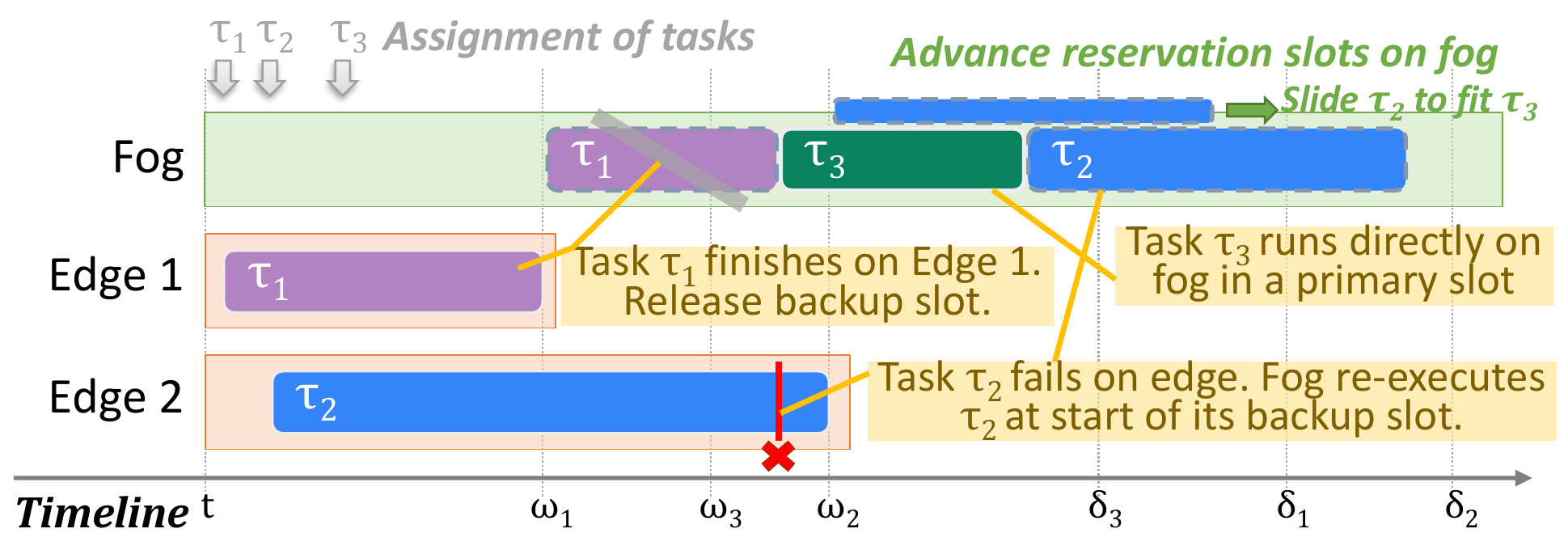}
	\caption{Task reservation and execution in a Fog partition}
	\label{fig:task-exec}
\end{figure}

We use the concept of \emph{advanced slot reservation} on the fog to provide these guarantees~\cite{ramakrishnan2009vgrads}. Slots are \emph{slices of resource time} within the fog's future timeline that can be reserved for specific tasks to execute. Slots reserved for re-executing failed tasks from the edge are \emph{backup slots} that may or may not be used, while those reserved for running tasks directly on the fog are \emph{primary slots} that will be used.

Once the fog identifies the candidate edges for an inquiry to run task $\tau_y$, we sort them in increasing order of their $\kappa$ cost. Incrementally, for each edge, $\widehat{r}_j^E$, we check the parent fog for contiguous free slots of length $\frac{\theta_y}{\rho(r^F_i)}$, starting from the latest completion time on the edge, $\omega_j$, and till the sub-deadline $\sigma_y$. 
If so, the fog can re-execute the task on itself during this slot before the deadline even if the edge fails. We \textit{reserve} these slots as backup slots for the task, and return the edge and its cost to the Master as a viable option.

To check for such contiguous slots rapidly, we maintain the free slots of the fog in an \emph{interval tree} data structure and search for the \emph{worst-fit free slot} that is viable. Say $[t, t']$ is the relative start and end time intervals for the largest free slot. We test if $\max(\omega_{j}, t) + \frac{\theta_y}{\rho(r^F_i)} \le t'$ and $\max(\omega_{j}, t) + \frac{\theta_y}{\rho(r^F_i)} \le \sigma_y$, i.e., can we complete the task re-execution before the interval's end time and before the task's sub-deadline. The space complexity for the tree is $\mathcal{O}(n)$ where $n$ is the number of free or reserved slots, and the time complexity for most operations is $\mathcal{O}(\log{n})$.

If contiguous slots that meet this task's requirement are not available, the fog tries to \emph{reschedule existing slot reservations} to widen the contiguous free slots, while ensuring that the task sub-deadlines for these prior reservations are still met. Intuitively, we use a \emph{defragmentation heuristic} that attempts to consolidate consecutive reserved slots to enlarge the free slot. First, we select the largest available free slot, say, $[t,t']$, between $[\omega_y,\sigma_y]$ and slide the slot reservation for some task $\tau_z$ that immediately \emph{succeeds} it as far to the \emph{future} as possible, without violating its sub-deadline, $\sigma_z$, or overlapping with its subsequent reservation. This will increase the $t'$ for the free slot, and hence widen the free interval, $(t' - t)$. If this slot is still not large enough for $\tau_y$, we attempt to slide the \emph{preceding} reserved slot for a task $\tau_w$ as far to the \emph{past} as possible without dropping below its edge completion time, $\omega_w$, or overlapping with its preceding reservation. This will decrease $t$ and widen the free slot. Lastly, we try to do both. This is repeated from the largest to the smallest free slot until one meets our needs.

If the slot reservation for a task is successful during an inquiry, the fog returns a viable bid to the Master. But this reservation is \emph{temporary}. If the \emph{bid is accepted} by the Master, it becomes permanent; if the \emph{bid is rejected}, the slot reservation is canceled. Also, if a task completes on an edge, its backup slot reservation is removed and available for future requests.

\para{Example} Fig \ref{fig:task-exec} illustrates this reservation process. Initially all the edges are free. When an inquiry for task $\tau_1$ comes with deadline $\delta_1$, the fog selects the cheapest free edge, \emph{Edge 1} for its execution from now till $\omega_{1}$. The fog reserves a \textit{backup slot} on itself from time range $[\omega_{1},\omega_{1}+t_1]$ for this task, where $t_1$ is the task execution duration on the fog. Similarly, task $\tau_2$ that arrives next is assigned to \emph{Edge 2} since Edge 1 is in use and Edge 2 is the next cheapest. It is initially reserved a backup slot on the fog from $[\omega_{2},\omega_{2}+t_2]$. Now when $\tau_3$ arrives, all edges are busy and the fog tries to reserve a \textit{primary slot} to directly run it. Since no slot is large enough to fit $\tau_3$ before its deadline $\delta_3$, we slide the slot for $\tau_2$ to the right but before its deadline $\delta_2$ to be able to allocate $\tau_3$ a slot. When $\tau_1$ completes execution on the edge, its backup slot on the fog is released.

\subsection{Reliable Execution of the Task} \label{sec:schedule:exec}
When a task is assigned to the fog after the Master accepts its bid, the parent fog makes the slot reservation for the task permanent. If required, the fog initiates transfer of the input \mb from the host resource to the edge the task will execute on; the fog caches a copy on itself for task re-execution or for direct execution. If the task is assigned to an edge, the fog invokes the task on the edge. If the task completes successfully by its deadline, the fog notifies the Master the output \mb ID and deletes the backup slot.

A \emph{timer} on the fog fires if we reach the start timestamp of a reserved slot. This can be a primary or a backup slot. If a primary, the fog invokes the relevant task on the cached input \mb to directly execute it within its sub-deadline and returns the output \mb ID to the Master. If a backup slot, the \textit{edge has failed to complete the task execution} on the \mb before its sub-deadline, either because the edge failed or due to under-performance -- a successful completion on the edge would have deleted this backup slot. The fog then re-executes this task using its cached input \mb, and responds back to the Master. Importantly, no active notification from the edge is required for re-execution if it fails.

\subsection{Scheduling Downstream Tasks}\label{sec:schedule:downstream}
The above steps are also applicable for successive tasks in a pipeline, with the difference being in how they are triggered. Once the source (or later) task completes execution, the edge, fog or cloud that executed it notifies the Master , and passes the output \mb ID and its location. This is equivalent to getting a trigger from the QE, except that we schedule the next task of an ongoing pipeline for the initial \mb. Note that multiple copies of a pipeline can be executing for the same DAG but on different input \mbs. Once the last task for a pipeline has executed before its deadline, the pipeline has completed; once all pipeline branches for a DAG have completed for an input \mb, the DAG execution has completed.

\subsection{Discussion} Using the fog for backup slots allows us to use cheaper edge resources reliably without paying for the fog resource if it is not used. If the edge capacities are all used up, we schedule directly on the fog by reserving primary slots at any point before the task's sub-deadline, to avoid going to the cloud. If edges are highly reliable, the backup slots are rarely used; but their existence guarantees task completion. But the fog cannot use these backup slots for directly executing other tasks and this can lead to under-utilization. We can extend the heuristic to allow over-subscription of fog slots during reservations by a certain ratio $\chi=\frac{\text{allowed load}}{\text{available capacity}} > 1$, i.e., reserve the same backup slots for up to $\chi$ tasks. The choice of $\chi$ can be decided based on the edge reliability or dynamic conditions. This allows the fog to be fully utilized, and avoids running tasks on the Cloud. But the trade-off is that it increases the chance of a failed task from an edge also failing on the fog if its over-allocated backup slot was used by another edge task that simultaneously failed. The $\chi$ ratio helps the user decide this trade-off.

\section{Experimental Evaluation} \label{sec:results}
\subsection{System Setup}
We use the VIoLET IoT emulation Environment~\cite{violet} to accurately mimic the behavior of an edge, fog and cloud deployment. 
VIoLET uses Docker containers hosted on VMs to replicate the compute and network performance of IoT resources.
We configure VIoLET with $111$ containers representing \emph{$100$ Raspberry Pi edges, $5$ Jetson TX1 fogs and $6$ 16-core cloud resources}, along with one cloud VM running the Master. Each fog partition has one fog and $15$--$25$ edges. These are hosted on $7$ Azure D32 VMs, each with a $32$-core Xeon CPU and $128GB$ RAM, and match the cumulative compute capacity of the $111$ resources. The relative compute performance ($\rho$) of edge, fog and cloud are $1$:$8$:$50$, based on real benchmarks.

The \emph{per core-hour billing cost} for the cloud is set realistically to $10$\textcent/hr, with the fog and edge being $10\%$ and $20\%$ cheaper, when normalized for performance. This gives the resource cost for edge as $0.167$\textcent/hr, Fog as $1.467$\textcent/hr. We use a $1~sec$ billing increment. The bandwidth and latency between Fog--Fog and Fog--Cloud is $100~Mbps$/$5~ms$, and for Fog--Edge as $60~Mbps$/$1~ms$. The bid timeout is set to $t_{inq}= 1~sec$ and the number of fogs to inquire is $n=2$.

\begin{figure}[t]
	\centering
	\subfloat[Task exec. time histogram]{
		\includegraphics[width=0.20\columnwidth] {figures//taskCount.pdf}
		\label{fig:nooftasks}
	}~
	\subfloat[Pipelines per DAG]{
		\includegraphics[width=0.20\columnwidth] {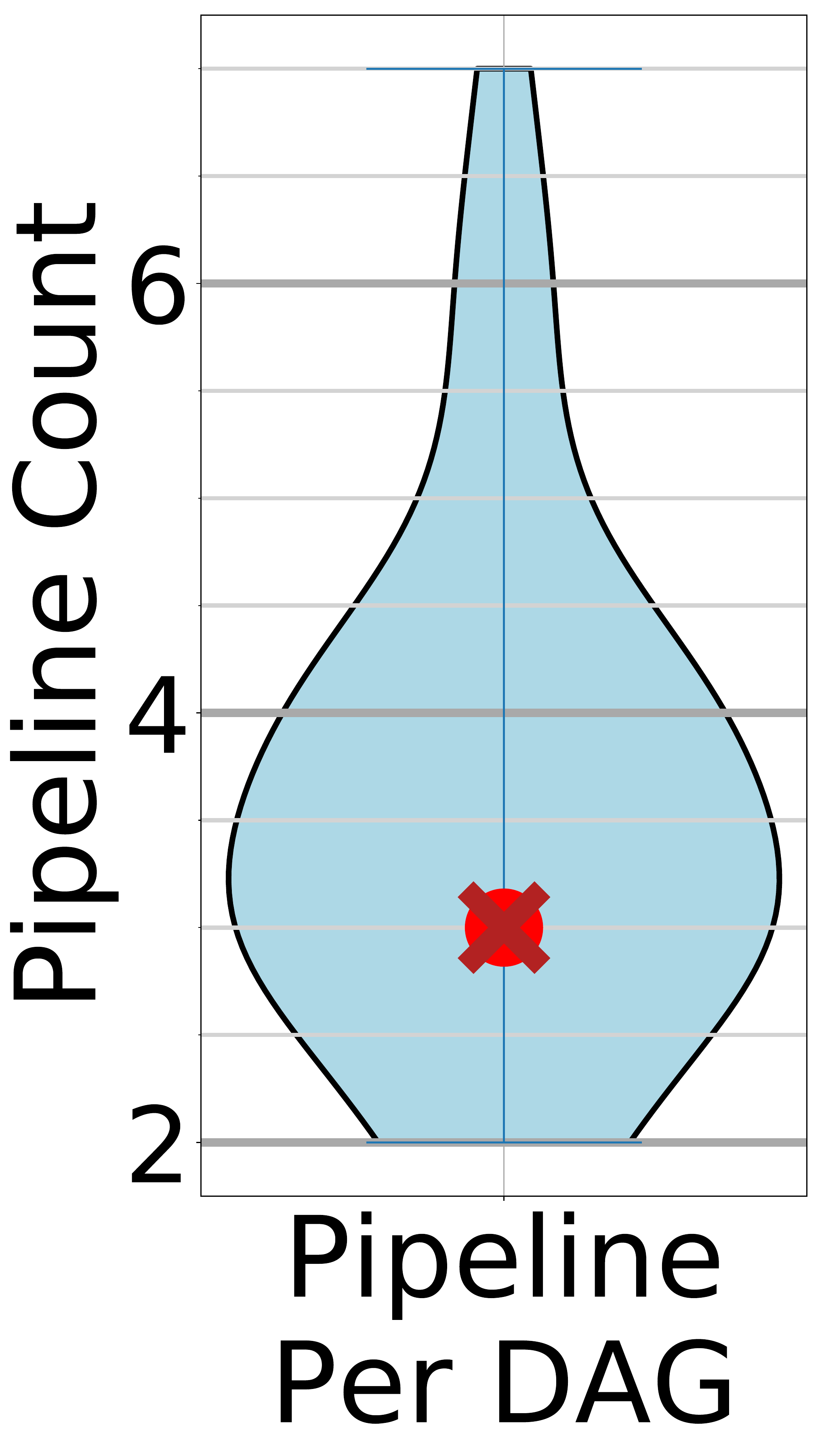}
		\label{fig:noofpipeline}
	}~
	\subfloat[Critical path time]{
		\includegraphics[width=0.20\columnwidth] {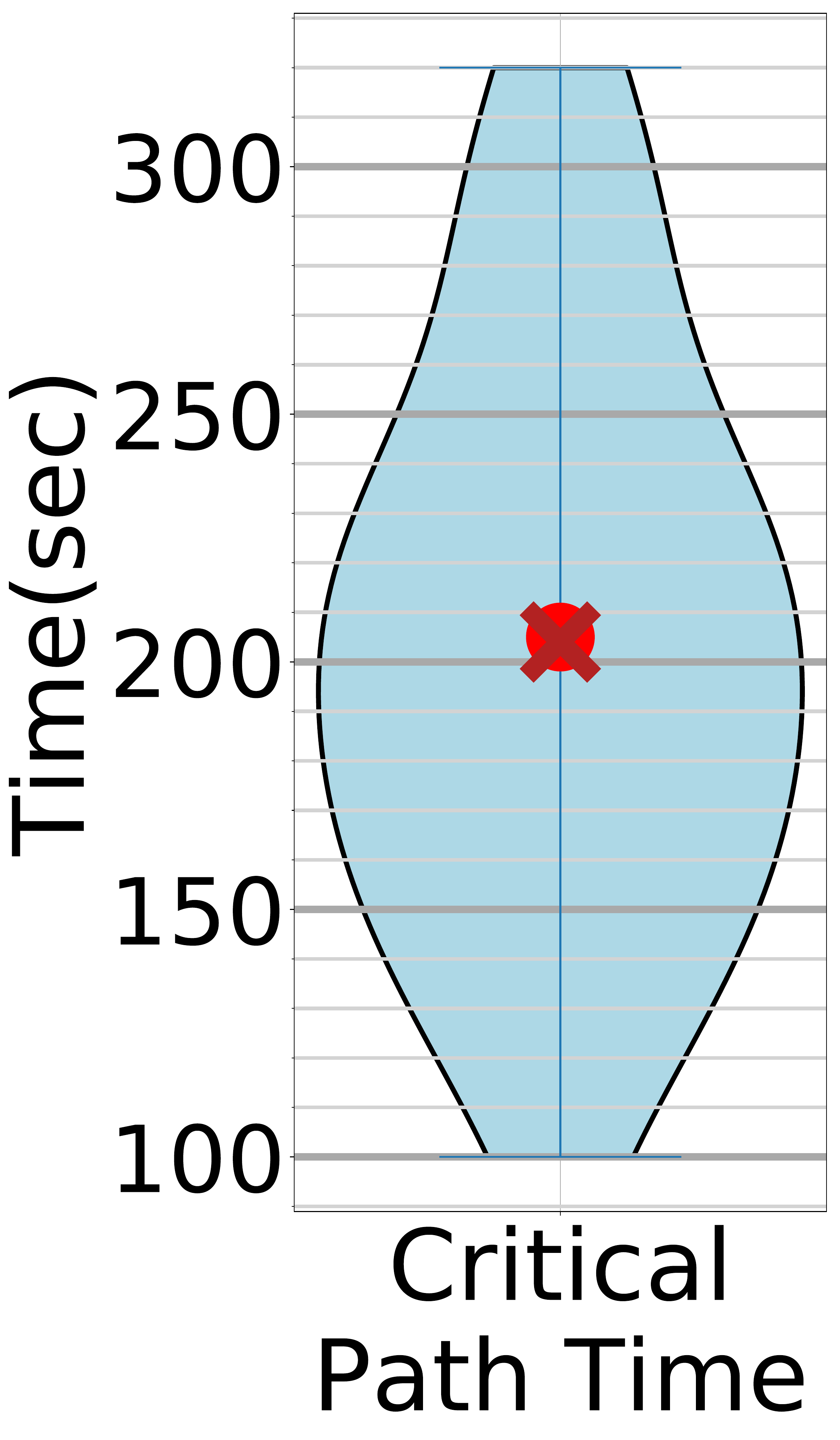}
		\label{fig:criticalpathtime}
	}~
	\subfloat[DAG exec. time]{
		\includegraphics[width=0.20\columnwidth] {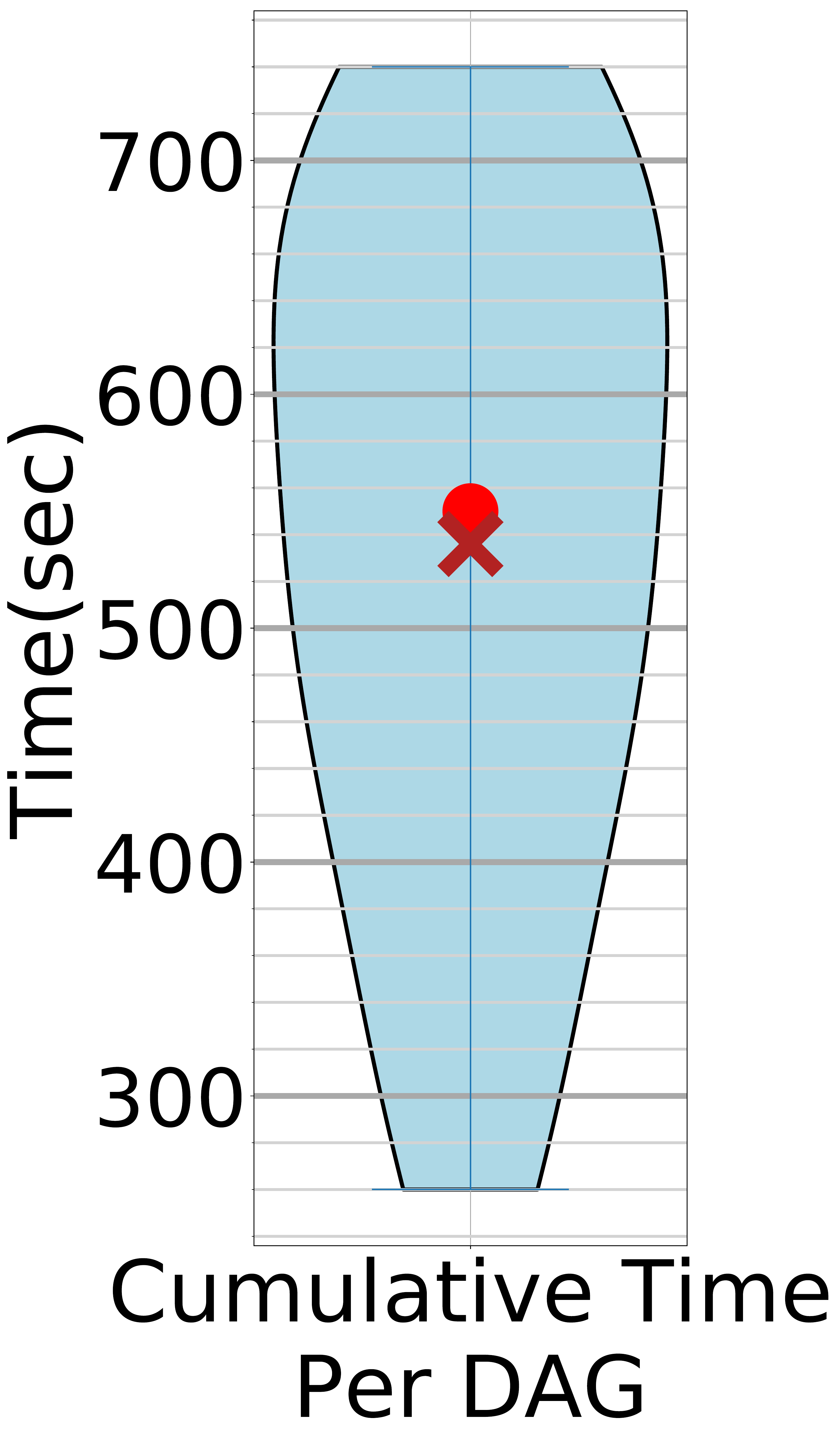}
		\label{fig:dagtime}
	}
	\caption{IoT Application Workload Characteristics.}
	\label{fig:exp:wl}
\end{figure}

\subsection{Application Workload}
We use $30$ DAGs sourced from permutations of the RIoTBench IoT dataflow workload~\cite{riotbench}. Each DAG has $2$--$7$ unrolled pipelines (Fig.~\ref{fig:noofpipeline}). We use $8$ synthetic tasks
with \textit{baseline execution time} of $\theta=10$--$60~$edge-secs, and their execution time on a resource linearly scales based on its performance scaling. These tasks are mapped to the RIoTBench DAG tasks with a Gaussian distribution (Fig.~\ref{fig:nooftasks}). The median critical path time for a DAG is $204~secs$ and the total time to execute all its tasks is $537~secs$ ($\approx 9~min$ to run a DAG on an edge (Figs.~\ref{fig:criticalpathtime},~\ref{fig:dagtime})). Synthetic \mbs with sizes uniformly from $500$--$1500~KB$ are generated from virtual sensors on edges. Their attributes match the filter query's ``hit-rate'' (\textit{\mbs triggered/sec}, and hence load on the system) for each experiment.

\subsection{Baseline Scheduling Strategies}
We compare our proposed scheduling strategy with three baseline algorithms to evaluate their costs and ability to complete the DAGs by the deadline.

\subsubsection{Cloud-Only (CO) Baseline} In this simple algorithm~\cite{saeid:fgcs:2013}, the edge and fog only generate \mbs from sensors and transfer them.
Every task that is triggered is executed only on the Cloud worker. Sufficient VM capacity is ensured to execute the peak number of concurrent tasks that are present. This makes it \emph{fast and reliable but at a higher cost}. The \mb still needs to move from its source resource to the Cloud. This is reflected in the execution time and the resource cost.
	
\subsubsection{Local Fog Partition (LFP) Baseline} This greedy algorithm executes all tasks in a DAG on the same fog partition containing the source \mb. Within the partition, it picks the cheapest available edge where the task can complete within its deadline. The parent fog is chosen only if all the edges are busy or cannot complete the task within the deadline, and the fog is free. There is no future slot reservations on the fog. A task (and pipeline) fails if the edge executing it fails or if inadequate resources are currently available in the fog partition for scheduling it. \emph{This localizes the scheduling decision and reduces the data movement to the partition, but may be unreliable as the task deadline is not efficiently used.}

\subsubsection{Fog Service Placement Problem (FSPP) Baseline} This adapts the deadline-aware application scheduling from~\cite{skarlat2017towards}. Tasks arriving with deadlines on a fog partition's queue are accumulated in a $1~sec$ window, solved as an Integer Linear Programming (ILP) scheduling problem using IBM CPLEX, and placed on local edges, the fog, the cloud, or a neighboring fog partition. Triggered Triggered DAGs are uniformly spread across the fogs, unrolled, and their tasks placed on the fog queue when ready. Only one task executes on a resource at a time, and the execution time on a resource depends on its performance scaling.

While \cf, CO and LFP actually trigger, schedule the DAGs and and execute their tasks for the \mb streams within VIoLET containers, we \textit{simulate} this workload for the FSSP scheduler and estimate the task completion and costs.

\subsection{Execution Cost Analysis with Reliable Edges} \label{sec:exp:cost}

\begin{figure}[t!]
		\centering
		\subfloat[Total Cost \textit{(left bar)} \& Avg. Cost for successful pipelines \textit{(right dot)}]{
			\includegraphics[width=0.3\columnwidth] {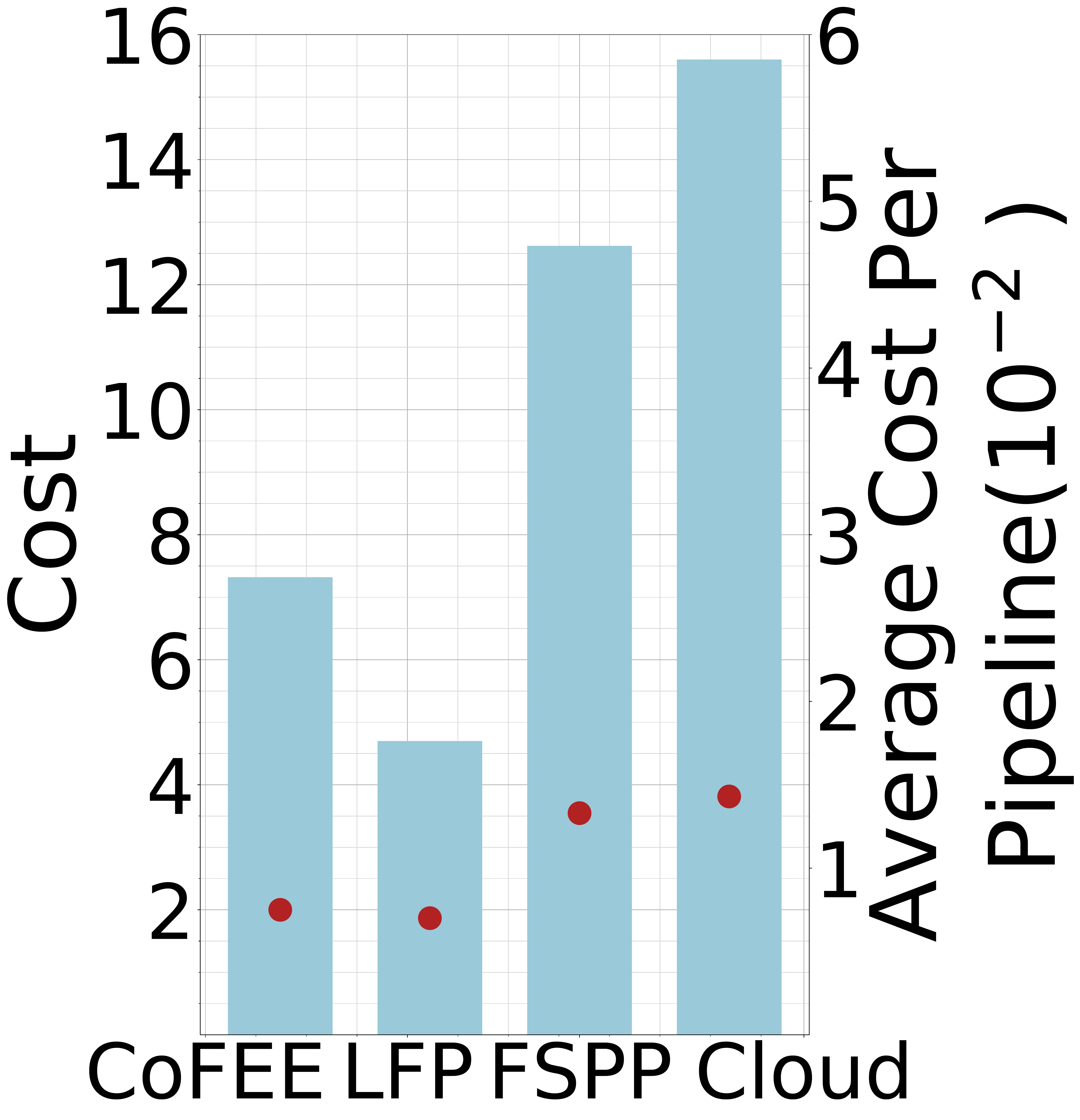}
			\label{fig:exp:cost:totcost}
		}~
		\subfloat[Pipeline success rate \textit{(green)} \& failure \textit{(red)} rate]{
			\includegraphics[width=0.3\columnwidth] {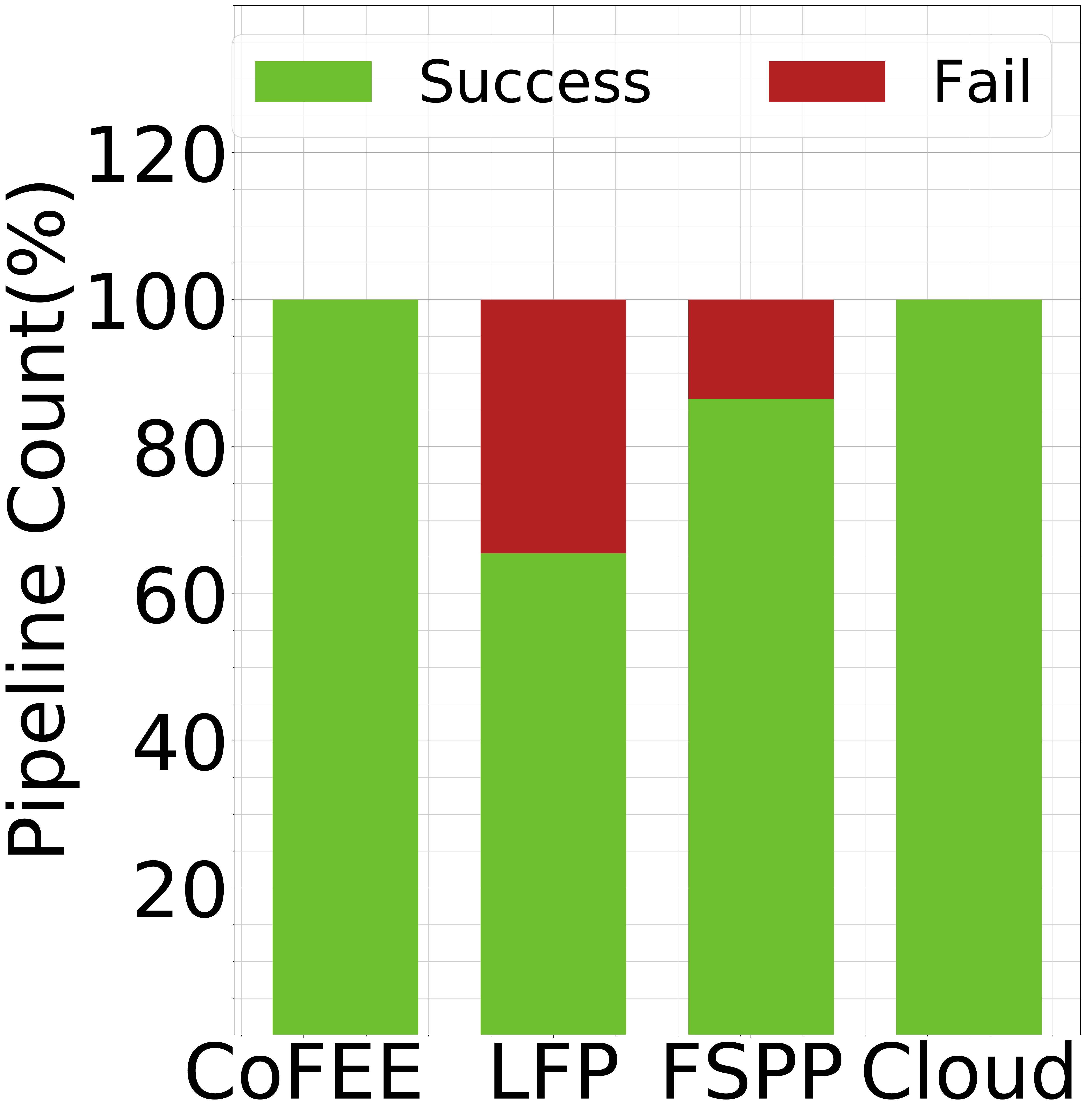}
			\label{fig:exp:cost:fails}
		}~
		\subfloat[Ratio of triggered tasks scheduled on \underline{E}dge, \underline{F}og \& \underline{C}loud)]{
			\includegraphics[width=0.3\columnwidth] {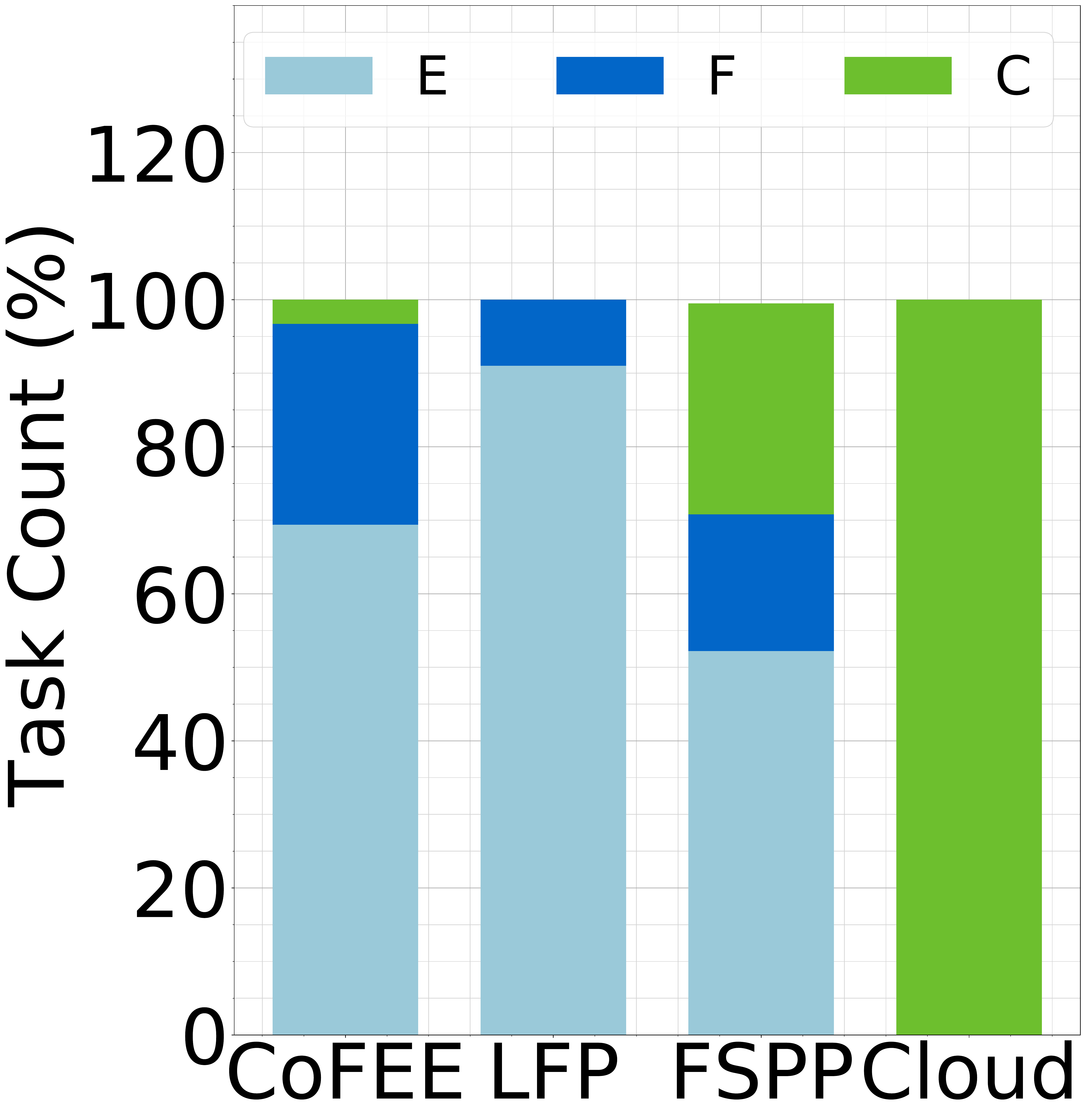}
			\label{fig:exp:cost:tasks}
		}% ~
		\caption{Performance of the schedulers with \textit{reliable} edges}
		\label{fig:exp:cost}
\end{figure}

%%%%%%%%%%%%%%%%%%%%%
In these experiments, we evaluate the ability of \cf to complete all the DAGs for \mbs they are triggered for, within the deadline and at a lower cost, compared to the baseline CO, LFP and FSPP schedulers.
We deploy the 30 DAGs on the 100 edge, 5 fog and 6 cloud workers with the default pricing. Here, we assume that all resources including edge are \emph{always reliable}. We configure the DAG filter queries such that the \emph{cumulative \mb triggering rate} is $15$~\textit{\mbs/min}, with a uniform probability on the edge source that generated the matching \mb and target DAG that it triggers. This translates to an \emph{average application load} of $100\%$ of the total edge and fog resource capacities, considering the basic scheduling overheads. 
We assign a \emph{deadline} that is $110\%$ of the critical path duration for each DAG when executing on edge resources, i.e., $10\%$ extra. We allow \cf to \emph{over-allocate} the slot reservations ($\chi > 1$) such that the \emph{full fog capacity} can be used for primary slots, and it can support backup slots for the \emph{cumulative edge capacity} in its partition. The experiments run for $20~mins$, and a total of $\approx 2740$ tasks are executed per run.

Figs.~\ref{fig:exp:cost} show the various metrics for \cf and the baselines. The total monetary cost for executing all the triggered DAGs is shown in Fig.~\ref{fig:exp:cost:totcost}, left Y axis bar. LFP has the smallest total cost, followed by \cf, FSPP and CO. LFP appears to be {$35\%$} cheaper than \cf.
However, when we consider the success rate in Fig.~\ref{fig:exp:cost:fails} that shows the fraction of pipelines that completed within the deadline, LFP a {$34.5\%$} failure rate while \cf and CO successfully complete all their tasks. This reduced load due to failures leads to a proportionally lower total cost for LFP. 
FSPP is $19\%$ cheaper than CO but almost twice as expensive as \cf, and exhibits a {$13.5\%$} failure rate. The cost per \emph{successful} pipeline execution (Fig.~\ref{fig:exp:cost:totcost}, right Y axis red dot) for \cf is similar to LFP and $50\%$ cheaper than CO and FSPP, with the latter two comparable.

Fig.~\ref{fig:exp:cost:tasks} shows the fraction of all triggered tasks that ran on each resource type for the three algorithms. \cf intelligently uses all three types -- edge, fog and cloud -- to offer $100\%$ completion within the deadline and at a low cost. Only $3.3\%$ of its tasks run on the cloud while $27.3\%$ on the fog and the rest $69.4\%$ on edge resources. As designed, the cloud worker is used only when the edge and fog do not have the capacity to complete the task within its deadline. There are no edge failures in this setup. Though the average count of DAGs triggered match the cumulative capacities of the edge and fog, the random generation of \mb causes load spikes which the cloud handles.
While LFP lowers costs by only using the local edge and fog, its reliability is lower by  not using the cloud, other fog partitions or future fog slots. Like \cf, FSPP also uses all three resource types but runs $28.7\%$ of the tasks on the cloud vs. only $3.3\%$ for \cf.

% %%%%%%%%%%%%%%%%%%%%%

\subsection{Application Resiliency Analysis with Unreliable Edges}
\label{sec:exp:fail}

\begin{figure}[t]
	\centering
	\subfloat[Total \textit{(bar)} and Average \textit{(dot)} pipeline cost]{
		\includegraphics[width=0.3\columnwidth] {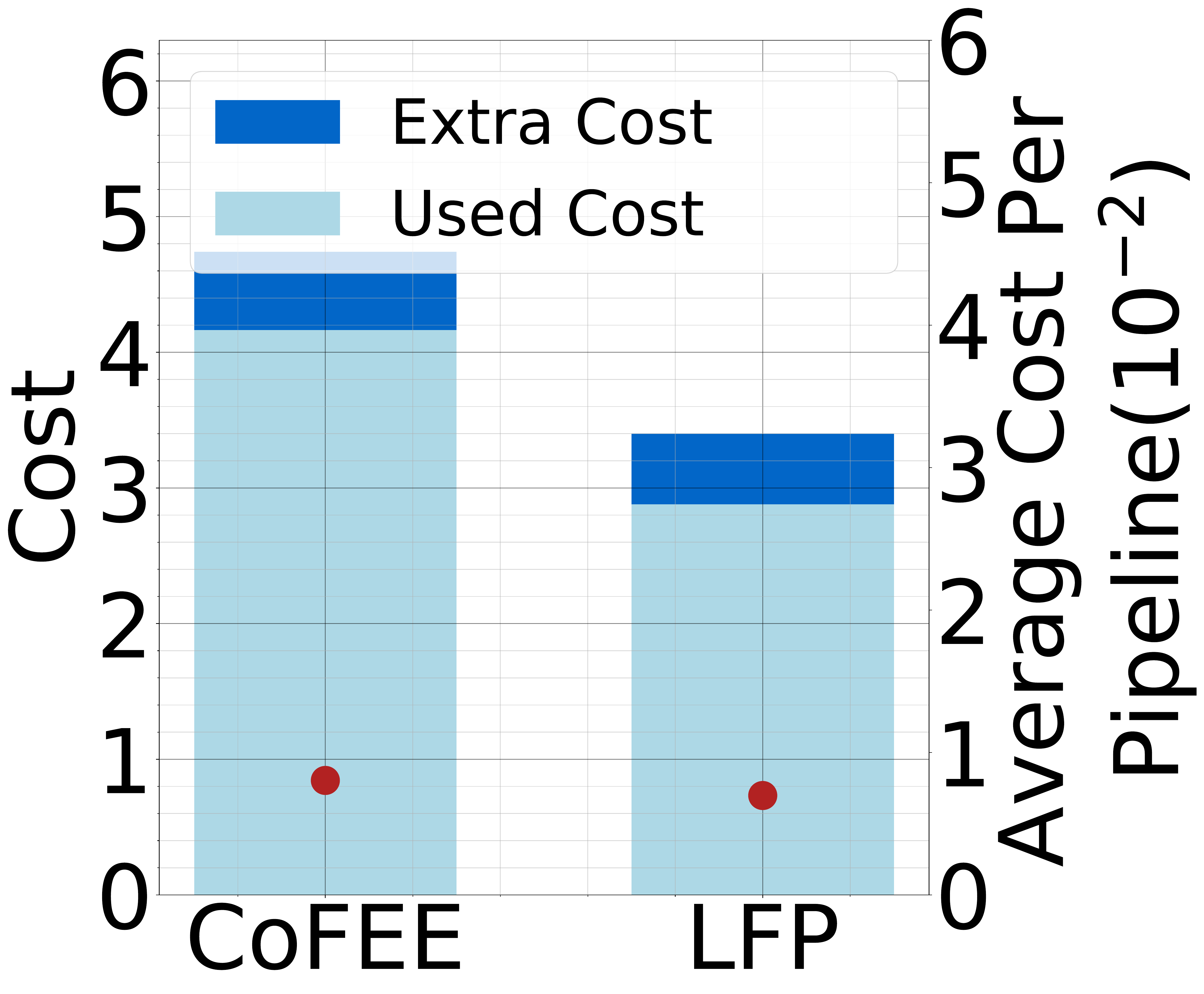}
		\label{fig:exp:fail:m100:totcost}
	}~
	\subfloat[Pipeline Success and Failure rates]{
		\includegraphics[width=0.3\columnwidth] {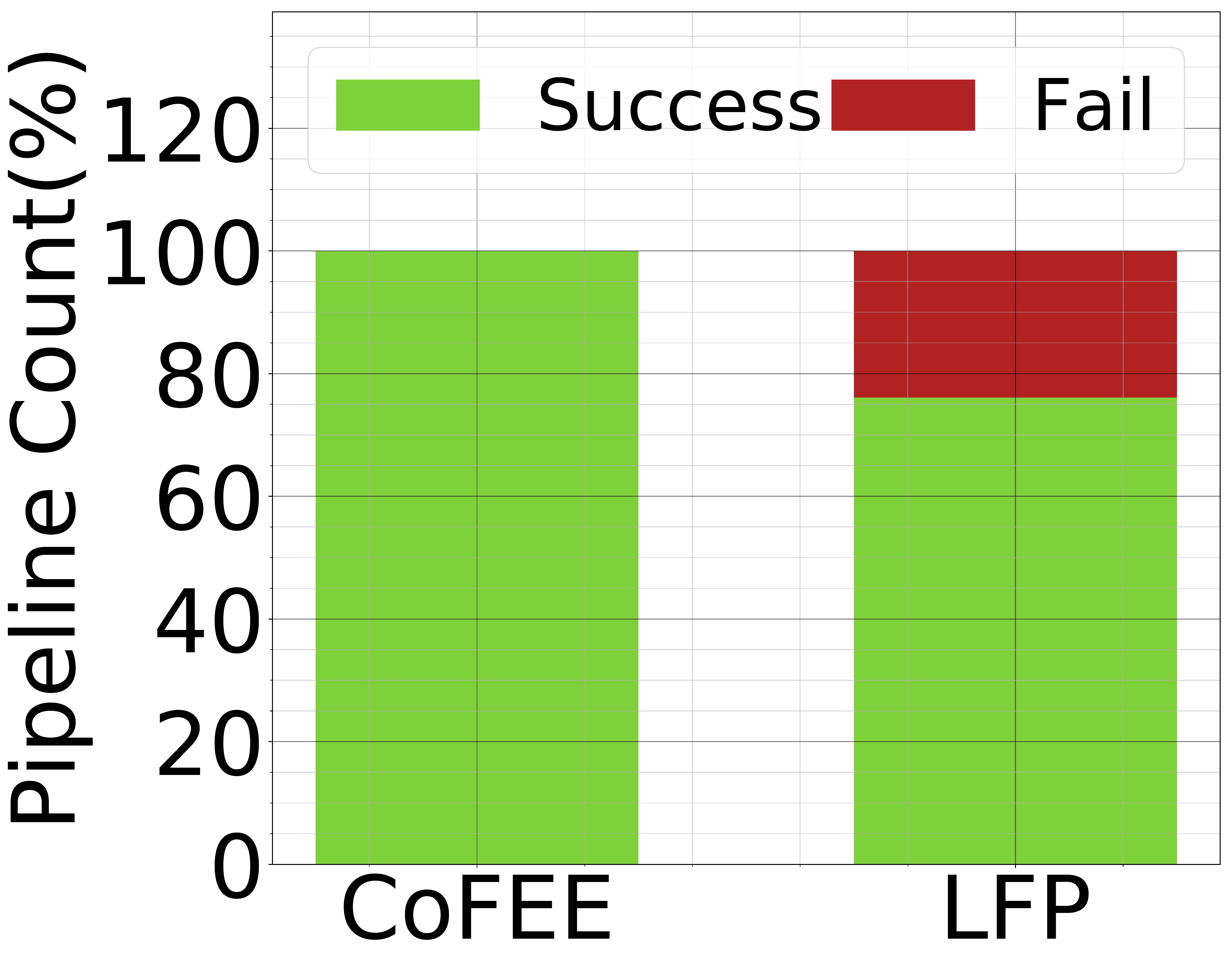}
		\label{fig:exp:fail:m100:fail}
		}~
        \subfloat[Tasks completed on \underline{E}dge, \underline{F}og \& \underline{C}loud]{
		\includegraphics[width=0.3\columnwidth] {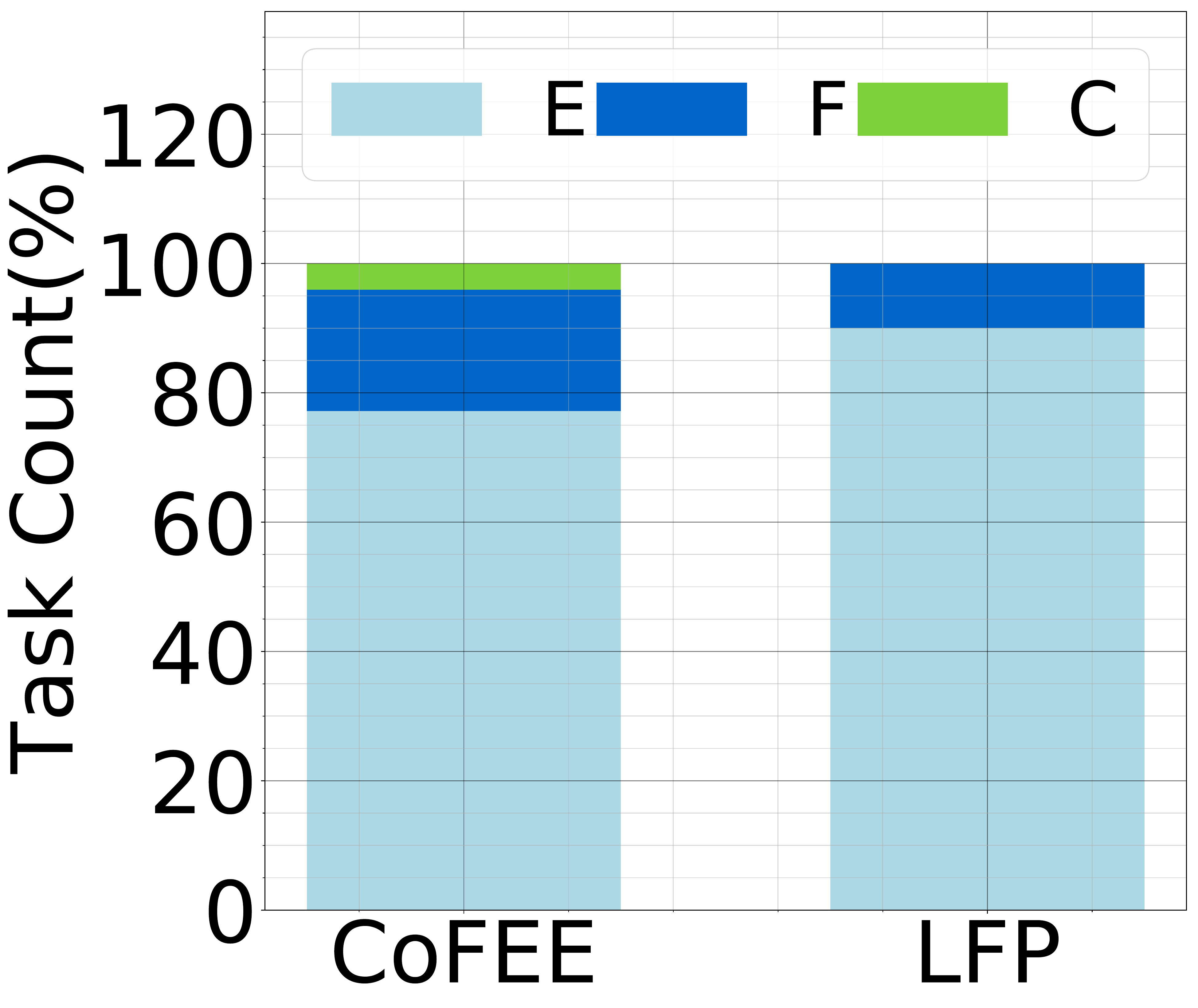}
		\label{fig:exp:fail:m100:tasks}
	}
	\\% ~
	\subfloat[Total \textit{(bar)} and Average \textit{(dot)} pipeline cost]{
		\includegraphics[width=0.3\columnwidth] {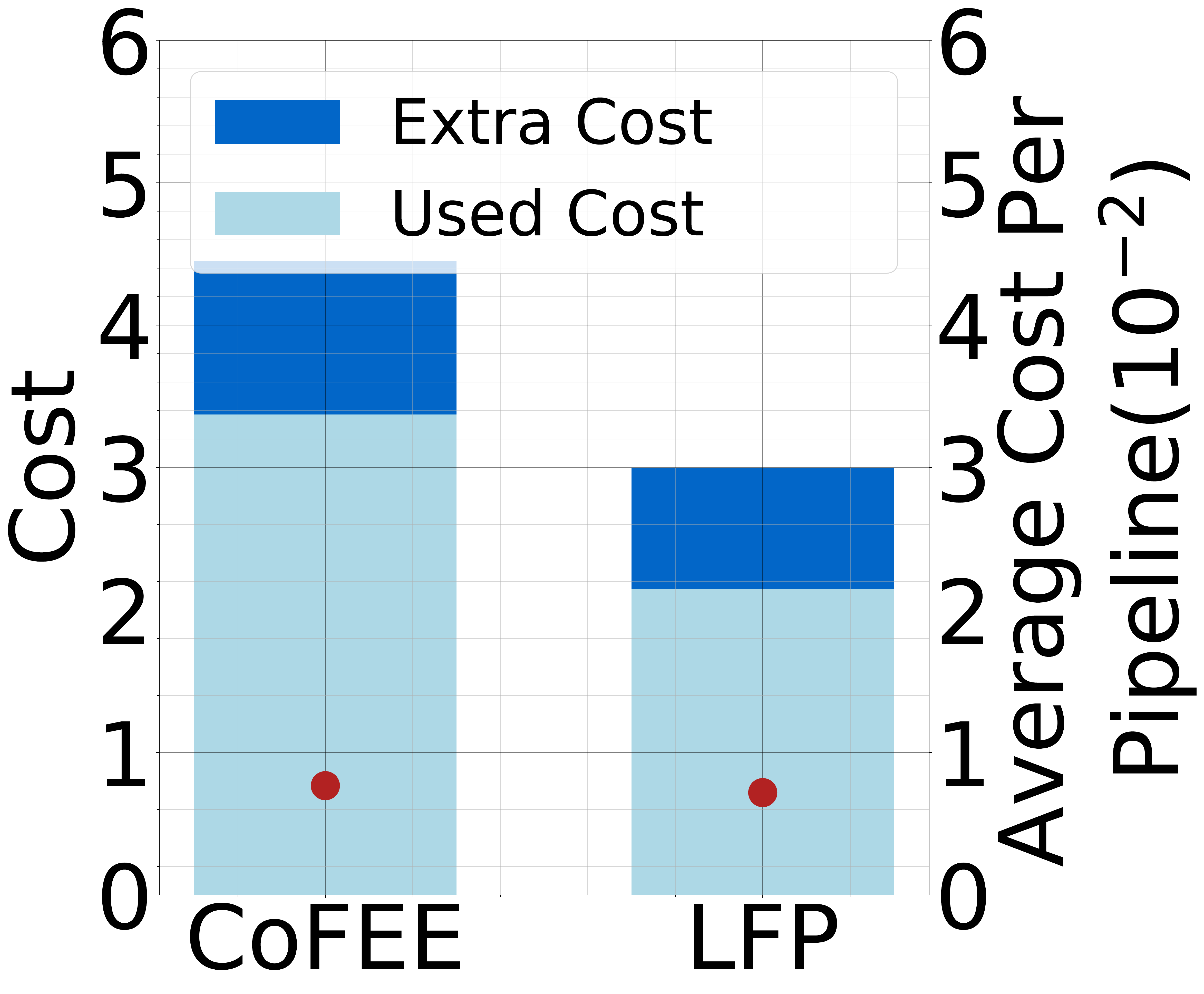}
		\label{fig:exp:fail:m40:totcost}
	}~
	\subfloat[Pipeline Success and Failure rates]{
		\includegraphics[width=0.3\columnwidth] {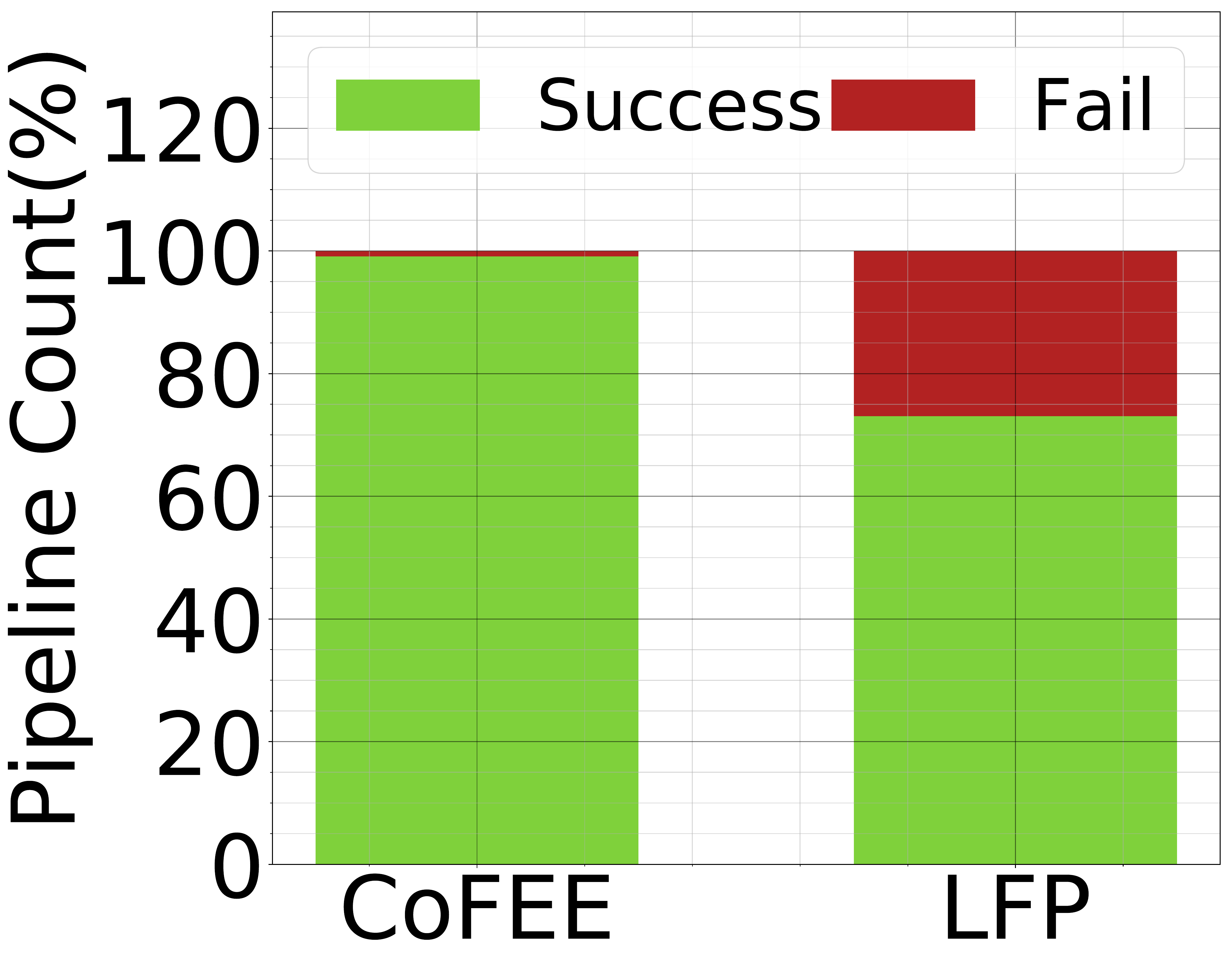}
		\label{fig:exp:fail:m40:fail}
	}~	
        \subfloat[Tasks completed on \underline{E}dge, \underline{F}og \& \underline{C}loud]{
		\includegraphics[width=0.3\columnwidth] {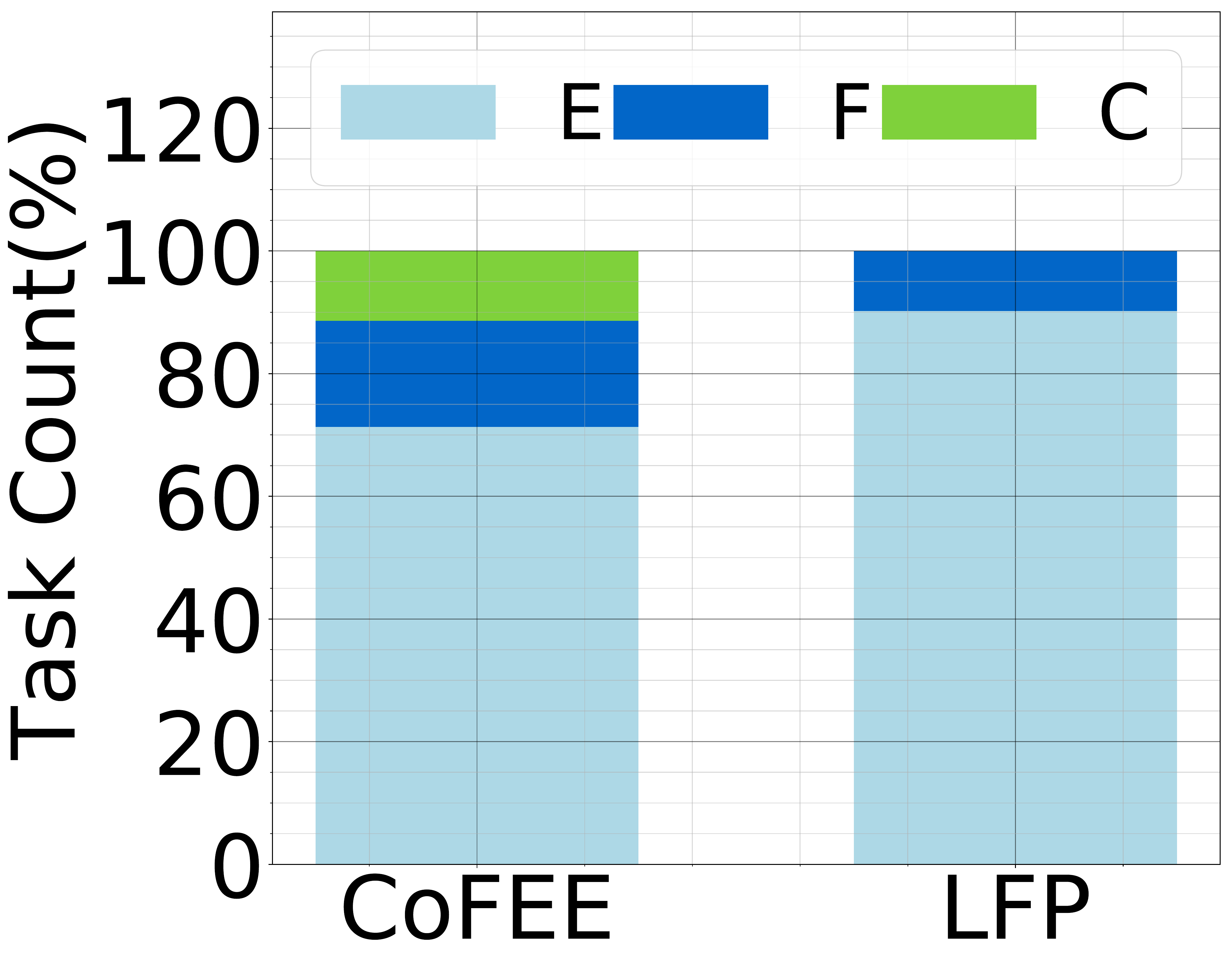}
		\label{fig:exp:fail:m40:tasks}
	}
	\caption{Performance of \cf and LFP on \textit{unreliable edges}, with MTBF of $100~mins$ (top row) and $40~mins$ (bottom row).}
	\label{fig:exp:fail}
\end{figure}

One of the key benefits of \cf is its ability to schedule applications with a high degree of resilience, even when the edge resources are fault-prone. Its advanced slot reservation heuristics on reliable fogs help achieve this. In these experiments, we evaluate the effectiveness of \cf in successfully completing DAG executions when edges have a \textit{low and high rate of failures}. We compare this against the LFP baseline; the CO baseline is not pertinent since it will always complete successfully on the cloud irrespective of the edge failures.

The experiment setup is similar to the above with some key differences. We use two configurations of MTBF for the Edge resources, $100~mins$ (M100) and $40~mins$ (M40). This means that with $100$ edges and an experiment runtime of $20~mins$, we expect $20$ edge failures ($\approx 1/min$) for M100 and $50$ edge failures ($\approx 2.5/min$) for M40. The edges fail \textit{independently} based on a uniform probability across time that matches these MTBFs. 
We have a mean application load that is $100\%$ of the cumulative capacities of the edge, which initially generates a cumulative of $10$~\textit{micro-batches/min}. This is smaller than earlier, where the application load could also saturate the fog for direct execution. Now, we expect to use the parent fog for backup slots that will be used due to edge failures. Since the edges are the source of the \mb, the loss of an edge proportionally reduces the application load and the compute capacity of the system, maintaining the average load at $\approx 100\%$ of the available edges. We retain a DAG deadline of $110\%$ of its critical path using only edge resources.

Fig.~\ref{fig:exp:fail} plots the various metrics for these two scenarios; M100 is on the top row and M40 on the bottom row. In Figs.~\ref{fig:exp:fail:m100:totcost} and~\ref{fig:exp:fail:m40:totcost}, we report the resource cost to execute tasks that are part of pipelines that eventually complete successfully (light blue) and the extra cost for tasks that execute for pipelines that finally failed and hence wasted (dark blue).

As before, LPF has a lower total cost for execution in both cases (Figs.~\ref{fig:exp:fail:m100:totcost} and~\ref{fig:exp:fail:m40:totcost}), explained by its large number of failures (Figs.~\ref{fig:exp:fail:m100:fail} and~\ref{fig:exp:fail:m40:fail}). 
The fraction of pipelines that fail for LFP has not increased from earlier even though edges fail in this setup. The failure rate for M40 is marginally higher than M100, $26.9\%$ vs. $23.9\%$, and both are lower than $34.5\%$ seen above. But unlike the previous setup where the application load was $100\%$ of both edge and fog capacities, now it is only at $100\%$ of the edge capacities. This lower load causes fewer failures. The failure rate difference between M100 and M40 is small. In its attempt to greedily schedule the tasks on a current free resource rather than use the deadline slack to defer the execution, LFP is often unable to find a free edge or fog. A small number of additional edge failures does not make this situation much worse. But the extra cost wasted by LFP on tasks of failed pipelines is higher for M40 than M100 since there is a greater chance of progress being lost for M40.

\cf is able to successfully complete \textit{all DAGs} in the M100 setup, and all but $5$ of the $580$ pipelines even in the M40 setup (Figs.~\ref{fig:exp:fail:m100:fail} and~\ref{fig:exp:fail:m40:fail}). In the latter, three of the failures happen because the fog was over-booked and executing a primary task during the backup slot, when the edges failed. Two failures happened because the edge was selected for execution but failed before the execution was started. 

We see from Figs.~\ref{fig:exp:fail:m100:tasks} and~\ref{fig:exp:fail:m40:tasks} that \cf makes greater use of cloud workers as the unreliable edges increase, running $4.1\%$ and $11.3\%$ of tasks on the cloud for M100 and M40. This partly explains the higher cost expended. As the failure rate increases, the edge cost spent on failed tasks also grows (Figs.~\ref{fig:exp:fail:m100:totcost} and~\ref{fig:exp:fail:m40:totcost}). $\approx 12\%$ and $\approx 24\%$ of the total cost is wasted for M100 and M40. However, unlike LFP, these tasks eventually succeed by re-executing on the fog.
That said, the mean cost per pipeline remains close to LFP that exclusively schedules on only edge and fog resources.

% %%%%%%%%%%%%%%%%%%%%%
\subsection{Scalability Experiments} \label{sec:exp:scale}

\begin{figure}[t]
	\centering
	\subfloat[Total \textit{(bar)} and Average \textit{(dot)} pipeline cost]{
	\includegraphics[width=0.4\columnwidth] {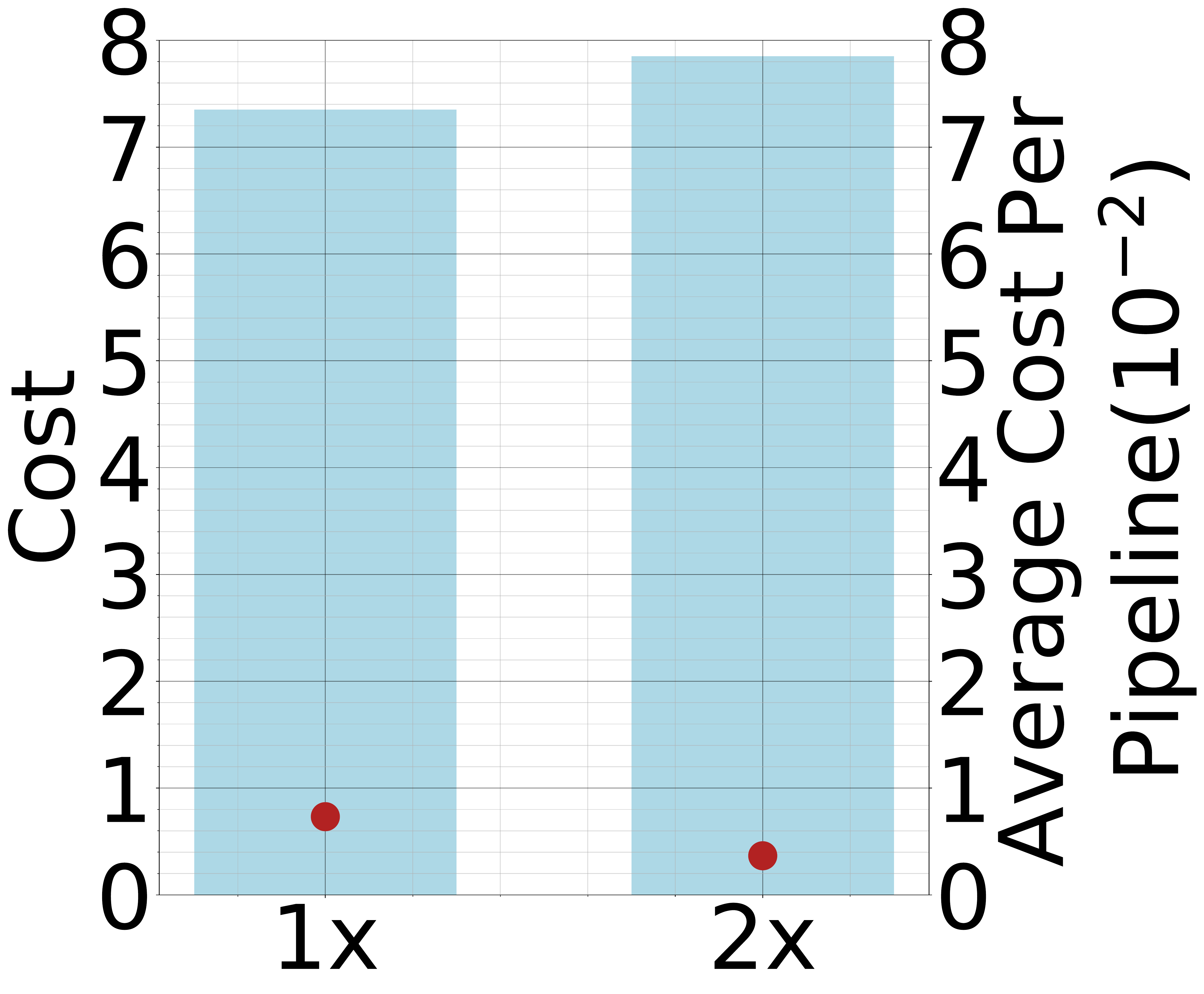}~~
		\label{fig:exp:scale:totcost}
	}~
	\subfloat[Task Throughput per Minute]{
		\includegraphics[width=0.4\columnwidth] {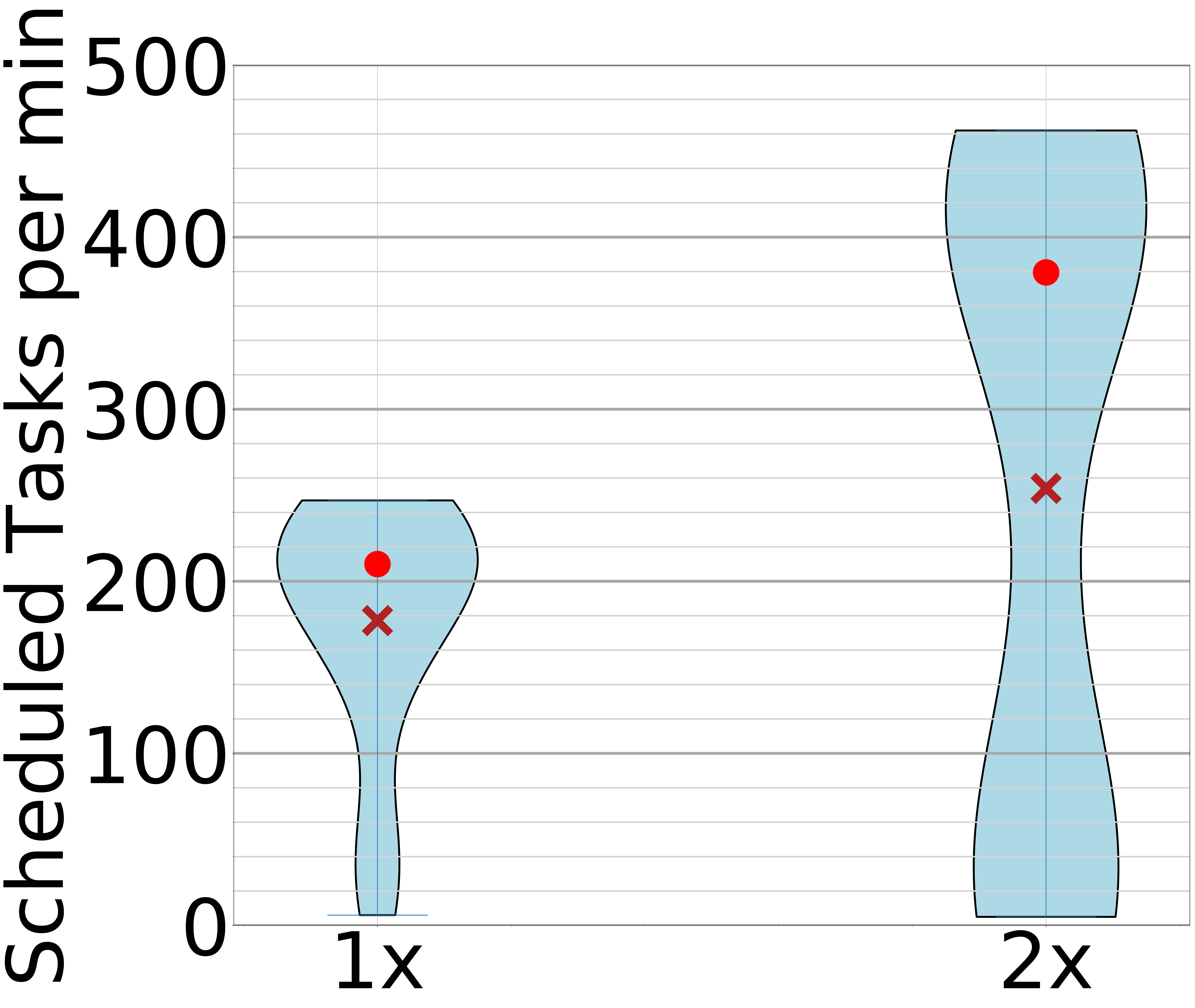}
		\label{fig:exp:scale:thruput}
	}
	\caption{Scalability of \cf scheduler and runtime.}
	\label{fig:exp:scale}
\end{figure}

Lastly, we examine the ability for the \cf runtime and scheduler to scale with the rate at which the DAGs that are triggered for execution. This reflects both the light-weight nature of our micro-service based implementation and also the limited runtime overheads for the scheduling algorithm.

We use a workload identical to the reliable edge experiments in Sec.\ref{sec:exp:cost}, which we call $1\times$WL. We then double the rate at which the \mbs are generated while halving the execution time per task in the $30$ DAGs. This maintains the application computation load on the cumulative edge and fog resources at $100\%$, and yet doubles the number of tasks that are scheduled per second. We also increase the DAG deadline to $125\%$ since some of the scheduling costs have a fixed time overhead. With shorter tasks, a larger fraction of the critical path time is used for these. This workload is called $2\times$WL.

Fig.~\ref{fig:exp:scale:totcost} plots the total cost for running these two workloads using \cf (bar on left Y axis). While the actual compute load for both these workloads are identical the $2\times$WL workload has a marginally higher total execution cost than the $1\times$WL workload. The key reason for this is the billing granularity of $1~sec$. For the smaller tasks, it is more likely that the billing round-up will cause excess payment for resources that are not used, and this accumulates across the $9154$ tasks that are executed for $2\times$WL. The average per-pipeline cost (red dot on right Y axis) is half for $2\times$WL as that of $1\times$WL. This is as expected as the tasks are half as long.
Fig.~\ref{fig:exp:scale:thruput} shows a violin plot distribution of the number of tasks scheduled per minute by the Master. The median task scheduling throughput for $2\times$WL is about twice that of $1\times$WL at $\approx 6~tasks/sec$. With higher rates of task generation, there is also more variability in the rate of tasks. 

In summary, the scalability of the system is limited by DAG triggering and task scheduling. QE can match $> 1000$ \mb-queries per second, per fog, and weakly-scales with the fog and stream counts. The $\approx 360$ tasks/min supported by our scheduler can improve by batching task scheduling requests to amortize the inquiry--bid--selection cost between cloud and fogs. These should allow us to scale to 1000s of streams.

% %%%%%%%%%%%%%%%%%%%%%
\section{Related Work} \label{sec:related}
There is an emerging body of work on scheduling of applications on the edge, fog and cloud ecosystem. Existing conceptual works deal with basic architectures, Application Programming Interfaces (APIs), and data communication~\cite{dasterdi:corr:2016}. In contrast, the question of how to effectively trigger, distribute and schedule streaming IoT applications across edge, fog and cloud resources has garnered less attention.

\emph{Publish-subscribe brokers} allow queries to be registered with topics. Published events that match the queries are routed to the subscriber~\cite{eugster2003many}. Brokers like MQTT are commonly used to route streams from sensors to application consumers through a well-defined topic for each sensor~\cite{mqtt}. But if every (large) \mb is sent as an event to a Cloud broker, it defeats the goal of moving compute to the data on the edge. If no queries match the \mb, the data movement has been unnecessary. If just the metadata is sent, we still spend the round-trip latency for triggering. Our application triggering design approaches a \textit{federated} publish-subscribe model, with the subscriber initiating the execution of a DAG. 
\emph{Complex Event Processing (CEP)} has been used for analytics over sensor events~\cite{triggered-iot-2016}. Here, we adopt it within the QE engine to match the \mb metadata and trigger DAGs.

\emph{Function as a Service (FaaS)} offer a ``serverless'' application model on the Cloud to execute simple stateless functions over input events~\cite{foxWorkshop}. Users also define a resource requirement. Invocations to these functions can then be seamlessly scaled by the provider on multiple VMs. FaaS are also offered on the edge by Azure IoT Edge~\cite{azure-edge}. Our stateless tasks resemble FaaS but offer dataflow composition, and leverage data locality and lower pricing when deciding where to run the function.

Research literature has examined \textit{programming models} for mobile edge, and on-demand fog and cloud resources along with a scheduling strategy~\cite{hong:mcc:2013}. Here, the applications are strictly wired as a tree, typically rooted in the cloud, and the tasks are deployed prescriptively at specific edge or fog hierarchy levels. We allow placement of tasks on any resource, based on deadline and cost optimization. They perform elastic resource acquisition when a fog is overloaded, and spatially repartition the process state, but do not consider the cost and deadline. \cf imposes a makespan deadline constraint with edge and fog having bounded capacities. 

There is a large body of work on scheduling applications on edge, fog and cloud~\cite{varshney2020characterizing}.
Some consider \textit{scheduling strategies} on fog and cloud~\cite{skarlatprovisioning} for a Bag of Tasks (BoT), where cloud resources are billed while the fog is free. They use fog to execute as many tasks as possible with two minimization goals -- the delay to propagate data to the cloud, and the cost. They extended this for scheduling DAGs with deadlines on fog and cloud using ILP~\cite{skarlat2017towards}, which we empirically compare against in Sec.~\ref{sec:results}. Here, if a fog partition is not able to accept a task, it off-loads it to a neighboring fog. Reliability is a non-goal. We instead consider a global view of all available resources and solicit bids to select the cheapest one. Our fog and edge resources are billed, albeit cheaper than the cloud.

Some papers use \textit{hard deadlines} -- more appropriate for mission-critical execution on streaming data -- and maximize resource usage on free fogs~\cite{yin2018tasks}. We use \mb and soft deadlines for tunable latency and throughput. Our edge and fog resources have usage costs, and the edge is unreliable. \emph{Energy constraints} are also considered in scheduling. Deng, et al.~\cite{deng2016optimal} investigate the power-delay trade-off as a metric for workload allocation in fog and cloud. They schedule transactional tasks on few fog and cloud resources to minimize their power usage while meeting the delay constraint. Others~\cite{brogi2017qos} map IoT tasks on Fog and Cloud resources using network bandwidth, latency and energy as constraints. In this paper, we consider energy to be subsumed by the pricing of the edge and fog.

There exists literature on distributed dataflow runtime and \textit{middleware} for edge and/or fog and/or cloud computing~\cite{ravindra2017mathbb,issarny2016revisiting}. However, they do not focus on application triggering and advanced scheduling. Some also consider migration of tasks and VMs for reliability and locality~\cite{bittencourt-2015}. We instead focus on light-weight task-level re-execution.

Several dimensions such as flexible dataflow triggering, hybrid resource reliability, and decentralized planning are missing from existing scheduling literature. We address these concerns here. We propose a novel declarative data-triggered approach for instantiating DAGs. These are scheduled on unreliable edge, and reliable fog and cloud resources to minimize the execution cost, and we use advanced slot reservations to meet the deadline and operate at scale. 

\section{Conclusions} \label{sec:conclude}

In this paper, we have presented a novel declarative model for matching \mbs generated by evolving stream sources with dataflows that are interested in consuming them, allowing developers to intuitively trigger their applications for sources based on their description ("what") rather than their endpoint ("where"). Our \mb model balances throughput against latency to scale to 1000s of streams. We have proposed a scheduling strategy to minimize execution costs on unreliable edges, and reliable fog and cloud resources using a federated bid--inquiry price discovery mechanism that scales with the resource count. It is also able to ensure reliable dataflow completion before its deadline using advanced slot reservations on the fogs. Our \cf platform implements these and is validated through comparative experiments on $30$ DAGs, $10k$ task runs and $>100$ edge, fog and cloud resources.

As future work, we plan to examine models that avoid complete unrolling of the DAG for efficiency. We also plan better recovery strategies based on keeping track of the execution lineage of tasks to avoid having to cache \mbs, and to support triggering over \mb generated in the past. More detailed experimental setups will be useful as well.

\section*{Acknowledgement}
This work was supported by grants from the Department of Science \& Technology, India under the Internet of Things (IoT) Research of Interdisciplinary Cyber Physical Systems (ICPS) Programme.
We thank the members of the DREAM:Lab at IISc, including Prashanthi S.K. and Deepsubhra Guha Roy, for their feedback on the CoFEE platform and the paper.

\bibliographystyle{plain}
\bibliography{paper}

\end{document}